\documentclass[fleqn,usenatbib]{mnras}

\usepackage{newtxtext,newtxmath}

\usepackage[T1]{fontenc}
\usepackage{ae,aecompl}

\usepackage{graphicx} 
\usepackage{amsmath} 
\usepackage{amssymb} 

\newcommand{\hh}{H$_2$}
\newcommand{\brg}{Br$\gamma$}

\newcommand{\Rb}{$R$(Br$\gamma$)}
\newcommand{\Rh}{$R$(H$_2$)}

\title[Effect of Slit Configuration on H$_2$~1-0~S(1)/Br$\gamma$]{The Effect of the Slit Configuration on the H$_2$~1-0~S(1) to Br$\gamma$ Line Ratio of Spatially Resolved Planetary Nebulae}

\author[I. Aleman]{Isabel Aleman$^{1}$\thanks{E-mail: bebel.aleman@gmail.com}
\\
$^{1}$UNIFEI, Instituto de F\'{i}sica e Qu\'{i}mica, Universidade Federal de Itajub\'{a}, Av. BPS 1303 Pinheirinho, 37500-903 Itajub\'{a}, MG, Brazil
}

\date{Accepted XXX. Received YYY; in original form ZZZ}

\pubyear{2020}

\begin{document}
\label{firstpage}
\pagerange{\pageref{firstpage}--\pageref{lastpage}}
\maketitle

\begin{abstract}
The \hh\,1-0~S(1)/\brg\ ratio (\Rb) is used in many studies of the molecular content in planetary nebulae (PNe). As these lines are produced in different regions, the slit configuration used in spectroscopic observations may have an important effect on their ratio. In this work, observations and numerical simulations are used to demonstrate and quantify such effect in PNe. The study aims to assist the interpretation of observations and their comparison to models. The analysis shows that observed \Rb\ ratios reach only values up to 0.3 when the slit encompasses the entire nebula. Values higher than that are only obtained when the slit covers a limited region around the \hh\ peak emission and the \brg\ emission is then minimised. The numerical simulations presented show that, when the effect of the slit configuration is taken into account, photoionization models can reproduce the whole range of observed \Rb\ in PNe, as well as the behaviour described above. The argument that shocks are needed to explain the higher values of \Rb\ is thus not valid. Therefore, this ratio is not a good indicator of the \hh\ excitation mechanism as suggested in the literature.
\end{abstract}



\section{Introduction} \label{sec:intro}

As the most abundant molecular species in planetary nebulae (PNe), the emission of \hh\ is of great interest. Molecular hydrogen lines have been detected in the near infrared (IR) spectrum of more than 130 planetary nebulae (PNe) \citep[e.g.,][see also Appendix~\ref{Appendix:A}]{treffers_etal_1976,Isaacman1984,kastner_etal_1996,Lumsden_etal_2001,Guerrero_etal_2000,davis_etal_2003,likkel_etal_2006,Ramos_Larios_2017,Gledhill2018}. \citet{kastner_etal_1996} found a \hh\ detection rate of 40 per cent. Recent deep imaging detections of \hh\ in small structures in PNe indicate that this rate may be even higher \citep{Fang_etal_2018, akras_etal_2017, Akras_etal_2020}.

Most of the published \hh\ spectroscopic observations of PNe uses narrow slits or small apertures\footnote{To simplify the text, only slit observations will be mentioned, but most of the discussion also applies or can be easily extended to observations with different shape apertures.} including only part of the planetary nebula (see Table \ref{tab:observations}). Sometimes only extractions of the slit observation are considered. Often the authors focus on specific positions in the nebula, for example, obtaining a spectrum with a slit centred at the \hh\,1-0~S(1) line emission peak. 

The position and width of the slit aperture during a observation have a great impact on the measured line fluxes and derived line ratios demonstrated by \citet{Fernandes_etal_2005}, \citet{Gesicki_etal_2016}, and \citet{Akras_Abell14} studies for optical atomic lines. \citet{Fernandes_etal_2005}, for example, studied the effects of the nebular area covered by the slit on the atomic line ratios and derived quantities in H~\textsc{II} regions. Their results showed that ratios of low to high-ionization lines are sensitive to this area. The ratios of [O~\textsc{II}], [N~\textsc{II}] and [S~\textsc{II}] optical lines to H$\beta$ can be affected by up to 30\% in relation to the ratio of the entire ionized nebula. The difference of the emitting regions of each of the lines (low-ionization lines are emitted in the outer layers of the nebula) is responsible for such high percentage. The effect of the slit configuration on atomic line ratios can be important when studying diagnostic diagrams or comparing observations to models as showed by \citet{Akras_Abell14}.

It is therefore expected that the effect of the slit aperture and position can also be important for the \hh\ line ratios to \brg\ as they are not produced in the same region. The \hh\ 1-0~S(1)/\brg\ ratio (hereafter \Rb) has been used in several studies of the molecular content of PNe, as proxies of the \hh\ quantity \citep{aleman_gruenwald_2004,aleman_gruenwald_2011} and assisting the diagnostic of the acting excitation mechanisms \citep{MarquezLugo_etal_2015}.

This paper studies the effect of the slit configuration on the \Rb\ ratio, using observations available in the literature and numerical simulations to verify and quantify such effect. The observations used here are described in Sect.~\ref{sec:observations} and a table is given in Appendix~\ref{Appendix:A}. The numerical simulations are described in Sect.~\ref{sec:models}. The results of the analysis of the slit configuration effect is presented in Section~\ref{sec:results}. Conclusions are summarised in Sect.~\ref{sec:conclusions}.

\section{Observations} \label{sec:observations}

For the present work, \hh\ and \brg\ line fluxes and ratios from observations were compiled from the literature. The data collected is presented in Table~\ref{tab:observations} (see Appendix~\ref{Appendix:A}). The table shows the \Rb\ line ratios obtained from spectroscopic observations of PNe and information on the corresponding slit configuration. The table also includes a few other relevant characteristics of the PNe, which are used in the present analysis.

\begin{figure}

\centering

\includegraphics[width=7.5cm,trim=1.50cm 1.60cm 0.9cm 0.8cm, clip]{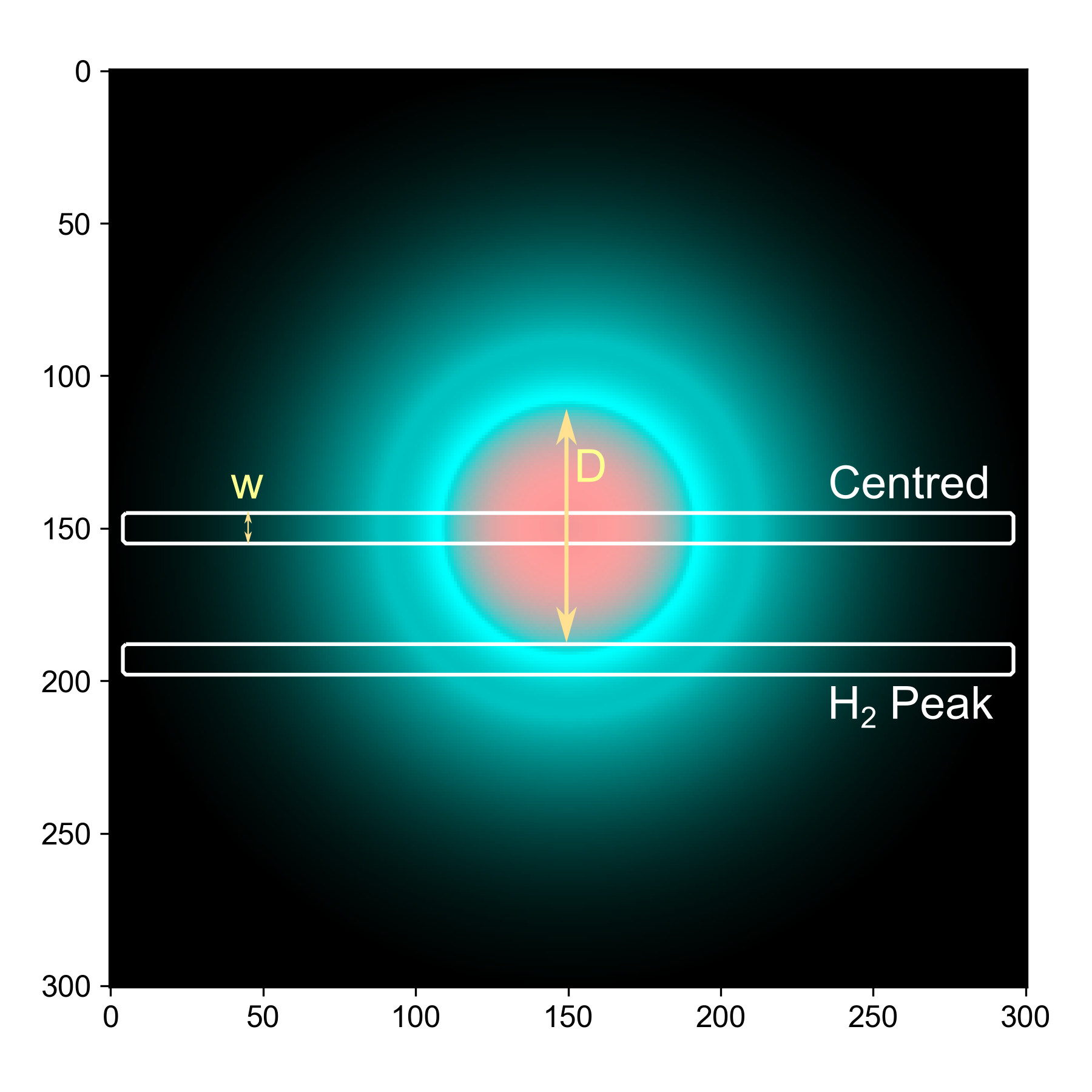}

\caption{Simulated image of a spherical PN in the \brg, \hh\,1-0~S(1), and \hh\,2-1~S(1) line emission, represented by red, green, and blue colours, respectively. The simulation uses the parameters of the reference model listed in Table~\ref{tab:stparam}. The two rectangles indicate the slit positions \textit{centred} and \textit{H$_2$ peak}. $D$ is the diameter of the PN and $w$ is the slit width.}

\label{fig:slitpositions}

\end{figure}

The slit centred at the nebular centre and positioned across the entire nebula (indicated as ``centred'' in Fig.~\ref{fig:slitpositions}) is the most typical PN observation configuration used for optical nebular analysis, i.e., for works focused on the ionized region. Although also used in studies of the \hh\ component in extended objects, other common position in this case is the narrow slit positioned at the \hh\ emission peak (indicated as ``H$_2$ peak'' in Fig~\ref{fig:slitpositions}). Observing the \textit{whole nebula} is only possible for more compact and/or distant PNe. In Table~\ref{tab:stparam}, the observations are classified within four general slit configurations as follows: 

\begin{itemize}

\item \textit{Centred}: the slit is positioned across the whole nebula passing by its centre;

\item \textit{\hh\ peak}: centred at the \hh\ 1-0~S(1) emission peak;

\item \textit{Whole nebula}: when the slit encompass the entire nebula;

\item \textit{Other}: all other configurations that do not falls under the previous categories or the configuration is not clear from the paper description.

\end{itemize}

\noindent This nomenclature will be used hereafter to indicate the slit positions for both observations and simulations. 

In Fig.~\ref{fig:slitpositions}, the illustration shows a round nebula, but the above classification is extended for all PNe morphologies. Bipolar PNe observations considered as \textit{centred} are those with the slit across the equatorial region, perpendicular to the symmetry axis. In bipolar PNe, the waist region is usually a bright structure in H$_2$ emission \citep{kastner_etal_1996}. This region shows a torus or barrel-like structure. For the purposes of this paper, what is of importance is the hydrogen species that are sampled by the slit (this will become clear further in the text.). It is then not difficult to see that using the slit across this torus/barrel is analogous to the \textit{centred} position for elliptical PNe, despite the angle the bipolar is observed. For bipolar PNe, observations considered as \textit{H$_2$ peak} are those where the slit samples a limited region around the wall of the torus, the wall of the lobes, or any known H$_2$ bright position.

\section{Simulations} \label{sec:models}

To simulate the slit observations of the nebula, we use the Python\footnote{\url{http://www.python.org}} library \textsc{PyCloudy} \citep[version 0.9.6;][]{Morisset_2013}. \textsc{PyCloudy} provides pseudo-3D simulations from \textsc{cloudy} one-dimensional models outputs.
The library produces simulated line emission maps, which allow the simulation of slit observations in any configuration and the calculation of the corresponding line fluxes.

Numerical calculations of the \hh\ and \brg\ emissivity in PNe were obtained with the photoionization code \textsc{Cloudy} \citep[version 17.01;][]{Ferland_etal_2017}. \textsc{Cloudy} can simulate a photoionized nebula from the ionized to the neutral regions (photodissociation region, PDR) self-consistently. As shown in \citet{aleman_gruenwald_2004,aleman_gruenwald_2011} and \citet{Aleman_etal_2011}, this is very important for the calculation of the \hh\ infrared emission, in special for the PNe with high-temperature central stars.

In the present models, the ionizing source emits as a blackbody, here described by its effective temperature ($T_\textrm{eff}$) and bolometric luminosity ($L_\star$). The gas is assumed to be spherically distributed. Models were calculated for two gas density ($n_\textrm{H}$) distributions: (i) uniform density and (ii) diffuse gas with a denser surrounding shell. For the first case, we studied density values from $n_\textrm{H} =$~10$^3$ to 10$^5$~cm$^{-3}$. In the second case, the diffuse gas is assumed to have $n_\textrm{H} =$~10$^3$~cm$^{-3}$ and the shell $n_\textrm{H} =$~10$^5$~cm$^{-3}$. Shell models were simulated with different central star temperatures and shell distances from the central star. For each set of central star parameters, simulations with four different shell distances were studied. The position were set where H$^0$/H = 10$^{-4}$, 10$^{-3}$, 10$^{-2}$, and 5$\times$10$^{-1}$. All simulations are done with solar abundance \citep{Grevesse_etal_2010}. Although different abundance sets may change the atomic line fluxes and therefore the gas cooling rates, such differences will not affect the conclusions of this paper. Dust is included in the models uniformly mixed with the gas. \citet{aleman_gruenwald_2004} showed that the dust composition in the regular dust model included in photoionization codes is not a significant factor for the \hh\ emission (due to the similarities in the general behaviour of their opacities). On the other hand, dust size and density may greatly influence the \hh\ emission \citep{aleman_gruenwald_2011}. Here, Cloudy models with graphite dust and dust grain size distribution typical for the ISM were explored. Dust-to-gas ratios of 3$\times$10$^{-2}$ and 3$\times$10$^{-3}$ were studied \citep{1999A&A...352..297S}. All the models were calculated assuming a 2~kpc distance, but this value has no effect on the present results, as they are based on distance-independent quantities. 

For convenience, reference values are defined in Table \ref{tab:stparam}. Unless stated otherwise the model parameters are the reference values listed in that table. 

\begin{table}
 \centering
 \caption{Reference Model Parameters.}
 \label{tab:stparam}
 \begin{tabular}{llcr}
  \hline
  Component & Parameter & Value\\
  \hline
  Central Star: & Temperature & 100\,000~K\\
  & Luminosity & 3\,000~$L_{\sun}$\\
  \\
  Gas: & $n_\textrm{H}$ & 1\,000~cm$^{-3}$ \\
  & Abundances & Solar\\ &&\citep{Grevesse_etal_2010}\\
  \\
  Dust: & Material & Graphite \\
  & Distribution & \textsc{Cloudy} ISM \\
  &&\citep{MRN1977}\\
  & Dust-to-Gas Ratio & 3$\times$10$^{-3}$\\
  \hline
 \end{tabular}
\end{table}

\begin{figure}

\centering

\includegraphics[width=\columnwidth]{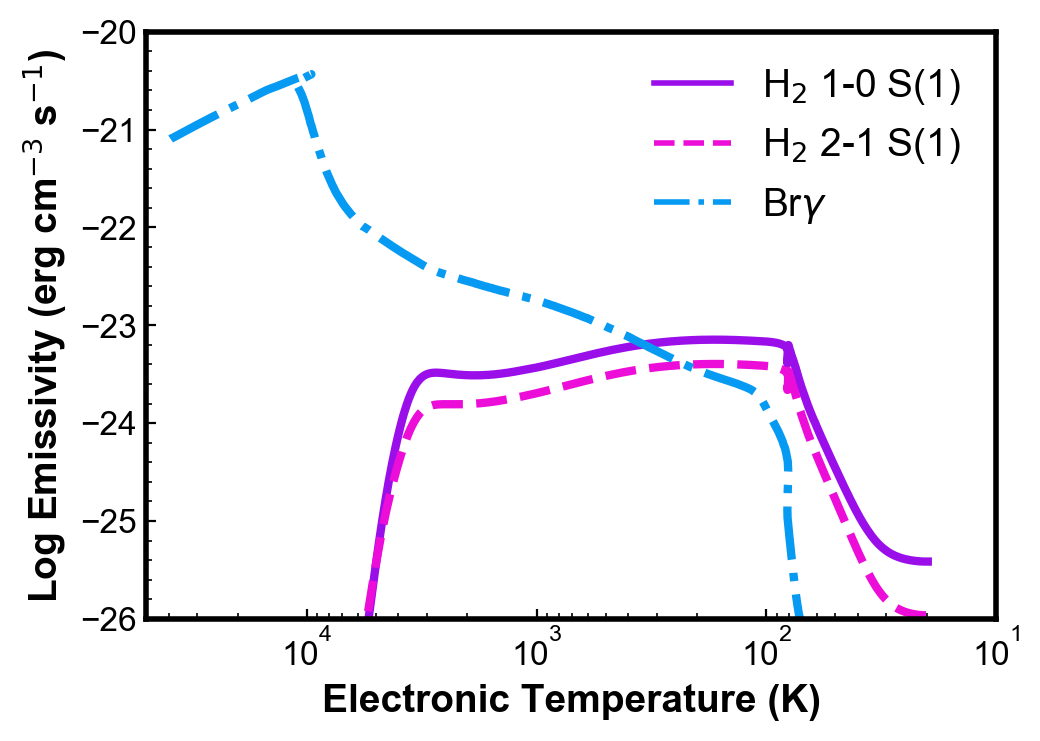}

\caption{Line emissivities as a function of the gas temperature for the reference model with parameters given in Table \ref{tab:stparam}. The gas temperature axis is shown in inverted order so the distance to the central star increases to the right.}

\label{fig:emiss_stop}

\end{figure}

The present models simulate the physical conditions and emissivities along the PN radial outward direction from an inner radius of 10$^{15}$~cm to a distance to the central star where the gas temperature decreases to $T_{\mathrm{F}} =$~40~K. This value was chosen based on inspection of the H$_2$ emissivity radial profiles calculated with Cloudy and to reproduce well the observations. Such stop criterion guarantee the inclusion of most of the \hh\ ro-vibrational lines 1-0~S(1) and 2-1~S(1) emitting region. Figure \ref{fig:emiss_stop} provides an example of behaviour of the emissivities of such \hh\ lines and \brg\ as a function of the gas temperature.

Different stopping temperatures were tested. For $T_{\mathrm{F}} >$~100~K (cut the nebula closer to the central star), a significant portion of the total nebular \hh\ emission would have been ignored (unless the cloud is limited by matter). For models with $T_{\mathrm{F}} =$~100~K, more than 70~\% of the flux determined for the model with $T_{\mathrm{F}} =$~40~K would have been ignored. For $T_{\mathrm{F}} =$~60~K, the flux ignored is approximately 50-70~\%. On the other hand, extending the nebula for very low $T_{\mathrm{F}}$, could produce an unrealistic large PN and the low line emissivity in such regions would not contribute much to the flux. For temperatures lower than 40~K, the \hh~1-0~S(1) and 2-1~S(1) line emissivities decreases and affects very little the total calculated flux (Fig.~\ref{fig:emiss_stop}). The difference in the fluxes found for $T_{\mathrm{F}} =$~40~K and $T_{\mathrm{F}} =$~20~K is less than a factor of two, which will not affect our conclusions.

The observation simulations were performed with {\sc PyCloudy}, using a slit placed in two positions, \textit{centred} and \textit{H$_2$ peak}, as discussed in the previous section. Fig.~\ref{fig:slitpositions} shows the configurations. The two slit apertures (white boxes) are positioned over a simulated PNe (reference model). The slit length is longer than the nebula size. For both cases, we study simulations where the slit width $w$ is varied from a small fraction of the nebula diameter up to a size that includes the whole nebula. The \textit{whole nebula} configuration can therefore be understood as a special limit case of both of the configurations above, when the slit width is large enough to cover the entire nebula.

\section{Results} \label{sec:results}

\subsection{The Effect of the Slit Configuration} \label{sec:sliteffect}

\begin{figure}

\includegraphics[width=\columnwidth,trim=0.3cm 0cm 0cm 0cm, clip]{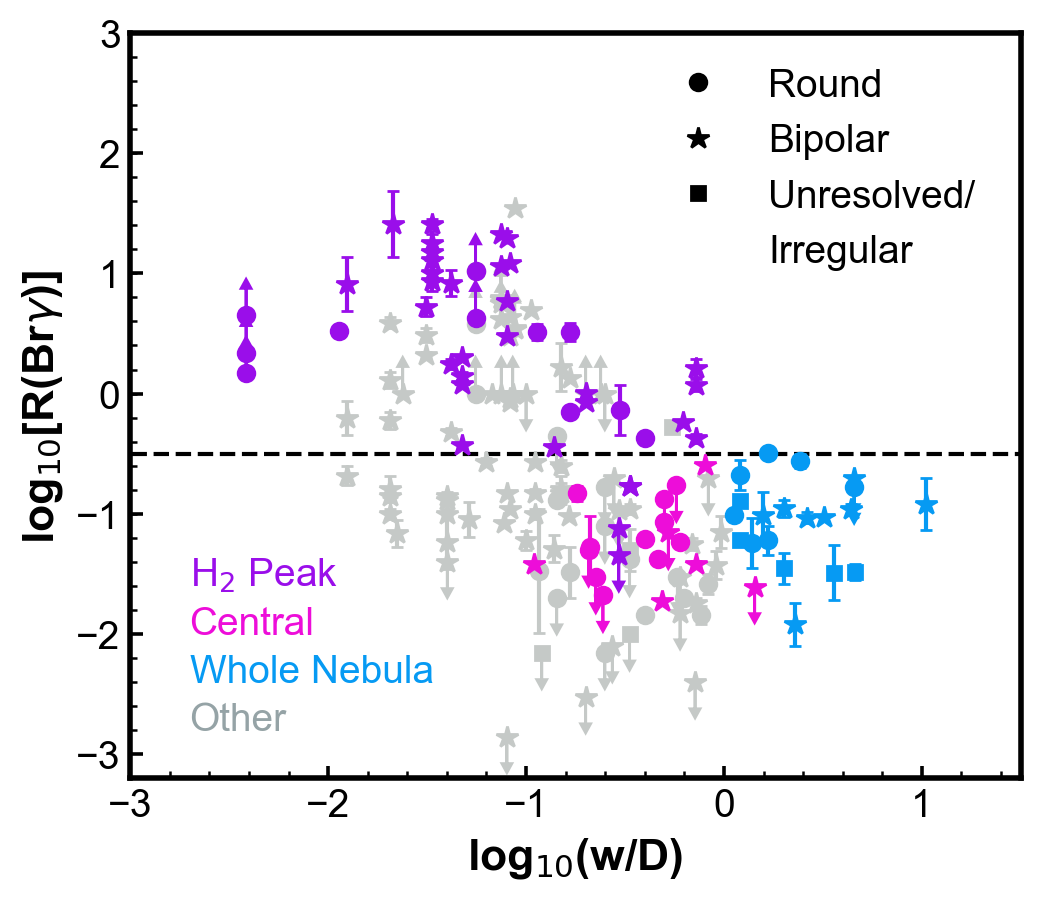}

\caption{\Rb\ as a function of the ratio of the slit width ($w$) to the diameter of the PN ionized region ($D$). Measurements are taken with configurations \hh\ peak (violet), centred (pink), whole nebula (cyan) and others (grey). The morphology of the object is indicated by the different symbols: dots for round, stars for bipolar and multipolar, squares for irregular and unresolved PNe. Error bars are shown when uncertainties are available. Upper and down arrows indicate lower and upper limits, respectively. The black dashed line indicates the maximum value found for whole nebula measurements.}

\label{fig:correlation}

\end{figure}

As the \hh\,1-0~S(1) and \brg\ lines are produced in different regions in the nebula (Fig.~\ref{fig:emiss_stop}) the effect of the slit configuration on the \Rb\ ratio in spatially-resolved observations can be significant. Indeed, Figure~\ref{fig:correlation} demonstrates that both the slit configuration and the fraction of the nebula covered by the slit can have a large influence in the \Rb\ values.

Figure~\ref{fig:correlation} shows \Rb\ as a function of the $w/D$ ratio, i.e., ratio of the slit width ($w$) to the diameter of the PN ionized region ($D$). The ionized region diameter $D$ is used for convenience, as it is more commonly available than the total PN size (including the neutral region). When this values is not found, other diameter is used. The value will be of similar order and as this only happens for a small fraction for the sample, it will not affect the results of this work. For bipolar PNe, $D$ is assumed to be the width taken in the minor axis of the object. As H$_2$ is more often seen in the torus/waist region or the wall of the lobes in bipolar PNe, the minor axis size would then be a better analogous dimension to $D$ of spherical PNe. The plot in Fig.~\ref{fig:correlation} includes all the data collected from the literature listed in Table~\ref{tab:observations}. The lower values of \Rb\ are limited for the observation sensitivity. This is what causes the empty lower left area in Fig.~\ref{fig:correlation}.

There is a clear segregation of \Rb\ values in Fig.~\ref{fig:correlation} related to the slit configuration. The ratios obtained in the \textit{centred} and \textit{whole nebula} configurations show a similar range of values. In these cases, there is no indication of significant trends \Rb\ $w/D$. The maximum value found for \Rb\ is 0.3 (indicated by the dashed line in the figure).

The \textit{\hh\ peak} and \textit{other} configurations exhibit the largest line ratios, which can be up to three orders of magnitude larger than the ratios measured for the \textit{whole nebula} or using the \textit{centred} configuration. For \textit{\hh\ peak}, most observation are found with \Rb~>~0.3.

There is also a trend for higher \Rb\ values being found towards small $w/D$. For \textit{\hh\ peak} observations, such trend can be naturally understood in terms of the different regions emitting \hh\ and \brg\ lines and the different regions covered by the slit in each configuration. For a given nebula, $D$ is fixed, so decreasing $w/D$, corresponds to using narrower slits. As $w$ decreases, the region around the \hh\ peak will include less \brg\ emission, maximising the ratio \Rb (see Figs.~\ref{fig:slitpositions} and \ref{fig:emiss_stop}). In the case of \textit{other} configuration, a similar behaviour might be occurring.

For the bipolar PNe observations, a classification scheme analogue to the scheme for round objects was used. Observations considered as \textit{centred} are those with the slit across the equatorial region, perpendicular to the symmetry axis. The torus is usually a bright structure in H$_2$ emission in bipolar PNe \citep{kastner_etal_1996}. Observations considered as \textit{H$_2$ peak} were taken with the slit sampling the wall of the torus, the wall of the lobes or at any known H$_2$ bright position. The results found, i.e. segregation and values, are the same as found for the round PNe.

\subsection{Simulations of \Rb\ for Different Slit Configurations} 
\label{sec:modelrep}

The study of models presented in this section has two goals: (i) to test our interpretation of the effect showed in the previous Section and (ii) to test if general photoionization models can reproduce the observed range of \Rb\ values if the slit configuration is taken into account. The second goal is of interest, as it is sometimes argued that UV excitation simulations cannot reproduce the observed values of \Rb\ in PNe \citep[e.g.,][]{MarquezLugo_etal_2015}. Comparison of observed with modelled \hh\ emission in PNe are done using zero- or one-dimensional models in many published works \citep[e.g.,][]{1999A&A...342..823V,Lumsden_etal_2001,Kelly_Hrivnak_2005,aleman_gruenwald_2011,MarquezLugo_etal_2015,Akras_etal_2020}, but in most of them the slit configuration during the observation is not taken into account in the comparison. 

\begin{figure*}
\includegraphics[width=5.8cm]{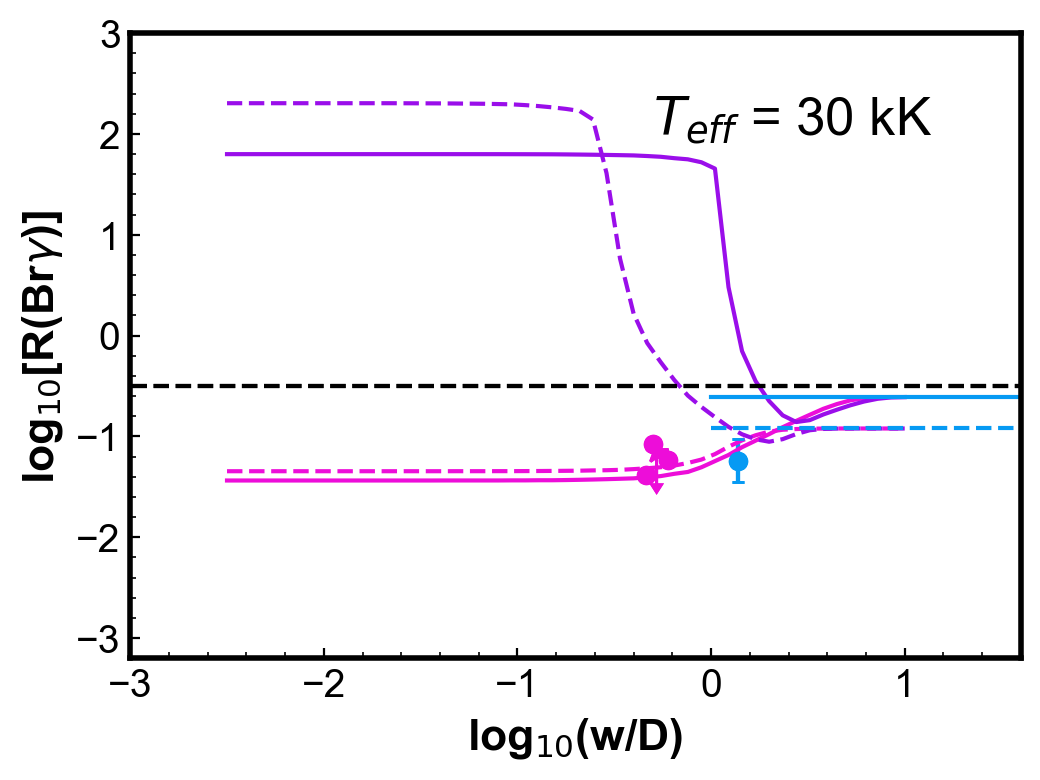}
\includegraphics[width=5.8cm]{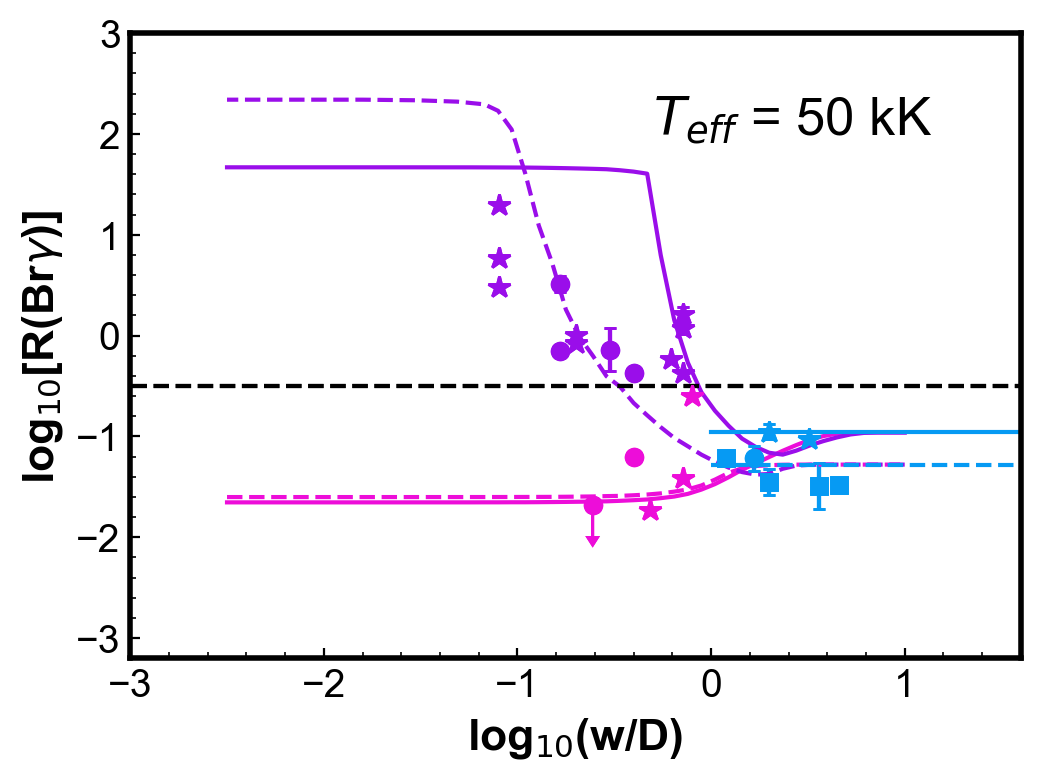}
\includegraphics[width=5.8cm]{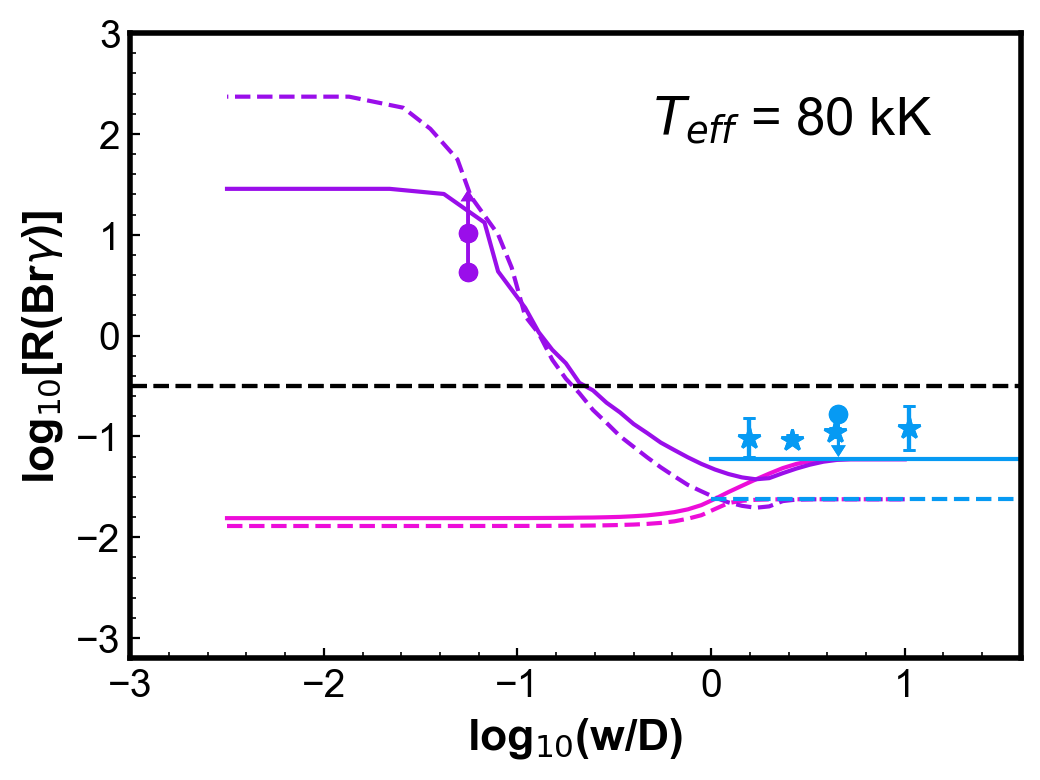}
\includegraphics[width=5.8cm]{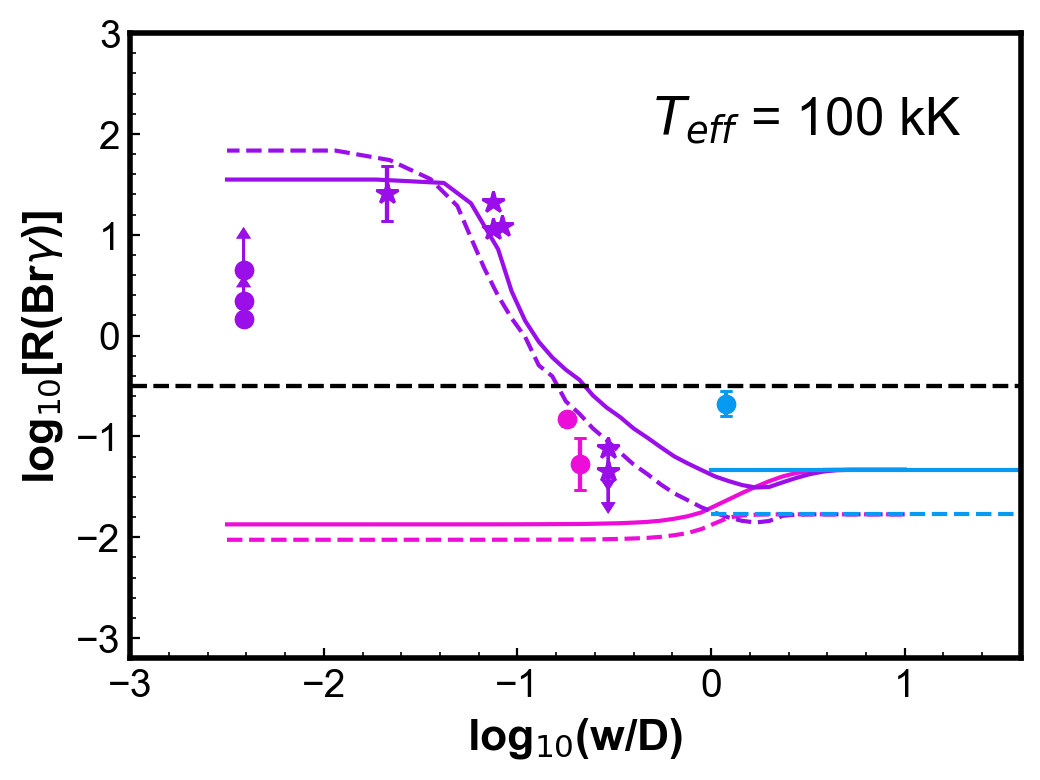}
\includegraphics[width=5.8cm]{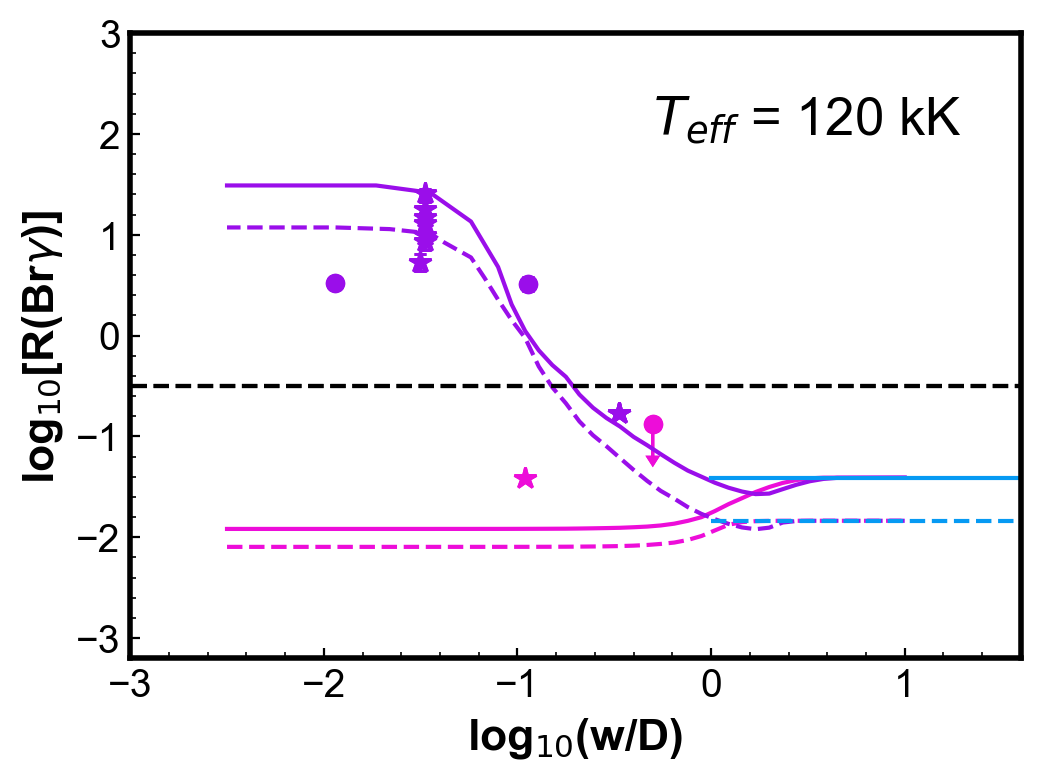}
\includegraphics[width=5.8cm]{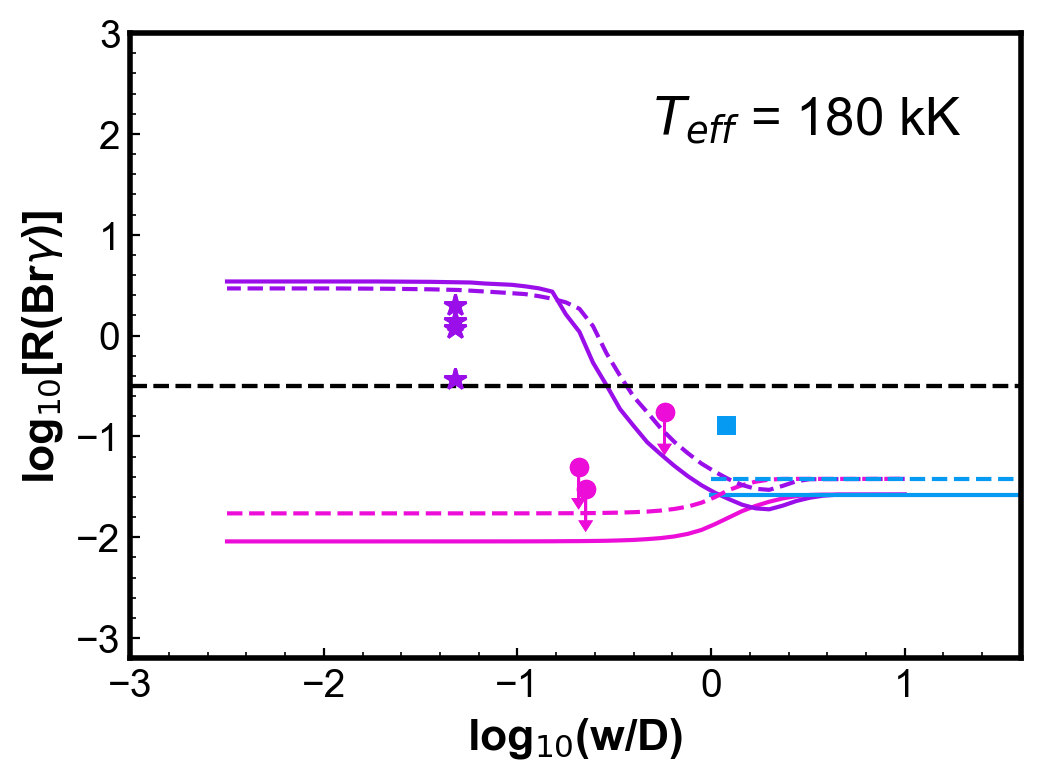}

\caption{Ratio \Rb\ as a function of the ratio $w/D$. Curves in each panel represent uniform density model with the $T_\textrm{eff}$ indicated. Solid curves uses dust-to-gas ratio of 3$\times$10$^{-3}$ and dashed curves 3$\times$10$^{-2}$. Dots are observations of PNe with a similar temperature ($\pm$~10~kK). The colour code indicates the slit position as in Fig~\ref{fig:correlation}. The black dashed line indicates the upper value found for the \textit{whole nebula} configuration.}

\label{fig:ModelsSphere}

\end{figure*}

Figure~\ref{fig:ModelsSphere} shows simulations of the \Rb\ ratio for round, uniform density PNe. Each panel shows two PN models calculated for a given $T_\textrm{eff}$ and with two different dust-to-gas ratios; the other parameters are from the reference model (Table~\ref{tab:stparam}).
For each PN model, three curves are generated: the cyan horizontal line indicates the values of \Rb\ calculated from fluxes integrated in the \textit{whole nebula}; the pink and the purple curves show the ratios calculated using the \textit{centre} and \textit{\hh\ peak} configurations, respectively, as a function of $w/D$. The colour code used for the models is thus the same as used for the observations in Fig.~\ref{fig:correlation}.

Observations of PNe with a similar temperature (within $\pm$~10~kK) are included in each of the plots for comparison. Even with the simplicity of these models, the calculated values can represent reasonably well the general behaviour and, in most cases, the magnitude of the observed \Rb.

For low $w/D$, ratios for the \textit{centre} position are smaller than for those obtained with the \textit{whole nebula} configuration. The difference is less than one order of magnitude. On the other hand, for the \textit{\hh\ peak} configuration, smaller values of $w/D$ produce larger values of \Rb, as the slit will be covering progressively less ionized emission, without varying much the molecular emission. A plateau is reached in progressively lower $w/D$ for increasing $T_\mathrm{eff}$. As seen in Fig~\ref{fig:emiss_stop}, even in the neutral region, there is a vestigial H ionization degree. For both configurations, when $w/D$ gets larger than one (slit covering most to all the nebula) the values tend to the \textit{whole nebula} value, as should be expected. 

\begin{figure}
\centering
\includegraphics[width=7.2cm]{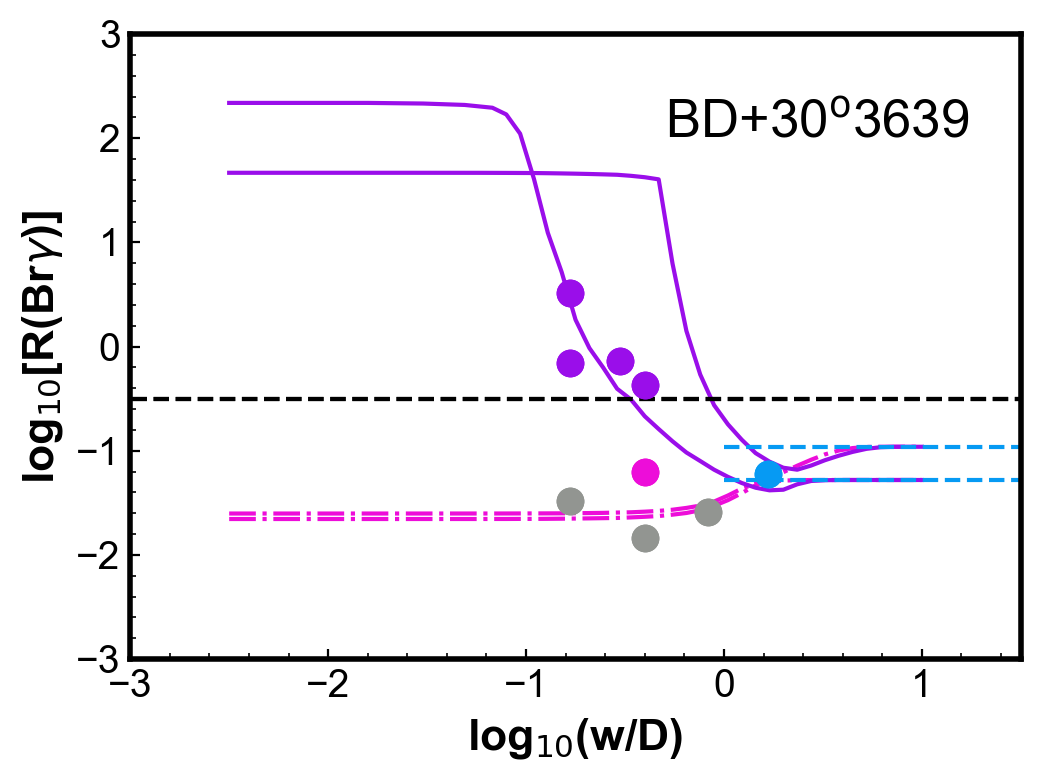}
\includegraphics[width=7.2cm]{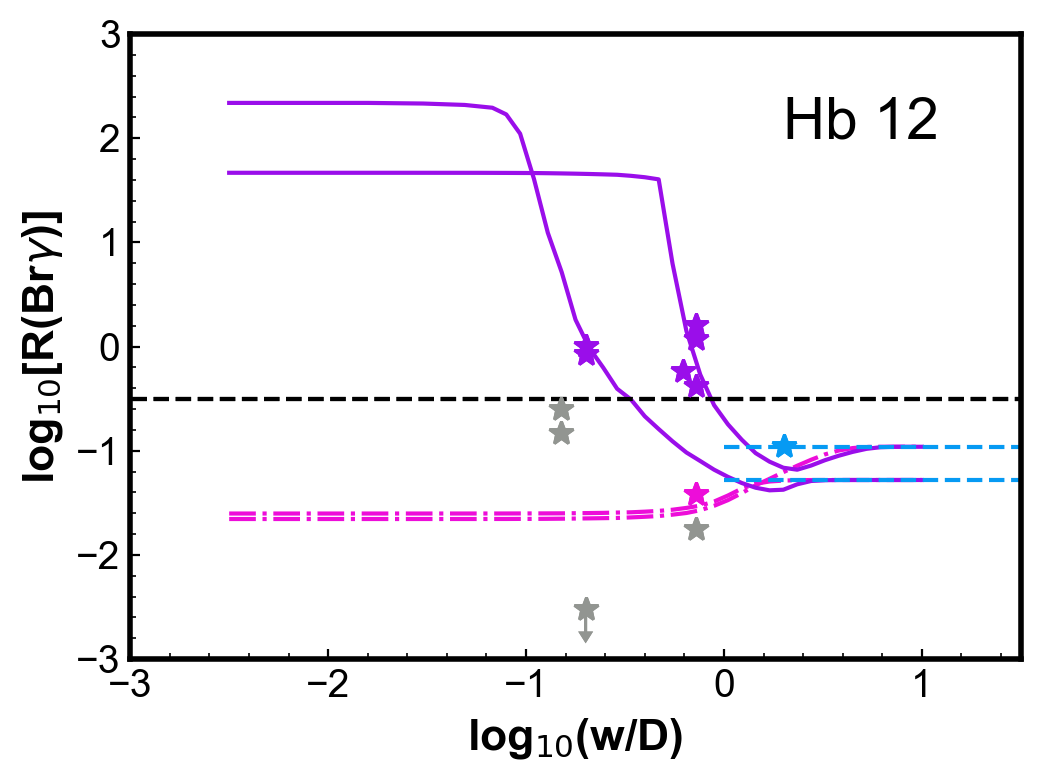}

\caption{Plots as in Fig.~\ref{fig:ModelsSphere}, where models included are for $T_{eff} =$~50~kK and observations are for the ring PN BD+30$^o$3639 in the top panel and for the bipolar PN Hb~12 in the bottom panel (in this panel error bars are not include as they, when available, are smaller than the marker size).}

\label{fig:BDandHb12}

\end{figure}

For two objects, BD+30$^o$3639 and Hubble~12 (Hb~12), there are a number of observations with different configurations published. For these two objects, there are observations using the all the configurations we discuss previously. The observation are shown individually in the plots of Fig.~\ref{fig:BDandHb12}. The models from Fig~\ref{fig:ModelsSphere} with $T_{eff} =$~50~kK, which is close to the object's $T_{eff}$, are also included. The models presented in this section were not developed to fit specific objects. No attempt to match specific characteristics of the object (apart for the close $T_{eff}$) was done. As mentioned above, these are simplified models. The goal here is only another sanity check, by verifying that the general behaviour and magnitude of the observed effect for individual objects are also reasonably reproduced by the models.

Models with different PN parameters, within typical ranges \citep[see][and references therein]{aleman_gruenwald_2011} were also studied. In Fig.~\ref{fig:ModelsSphere}, the differences in \Rb\ due the central star temperature and dust-to-gas ratio are shown. The general behaviour of the curves is similar for different star luminosities, within a typical PNe range. Changing the model luminosity in one order of magnitude for more or less change the curves in a similar amount as the change in dust-to-gas ratio shown. Variation in density within typical PNe values (10$^3$-10$^5$~cm$^{-3}$) may account for the differences between observations and models in \textit{whole nebula} and \textit{centre} configurations. However, models with high densities (>10$^5$~cm$^{-3}$) produces large differences between the \textit{\hh\ peak} \Rb\ calculated and observed. This is natural, as dense models produce very compact nebulae, while the \textit{\hh\ peak} observations usually probe sub-structures in more extended PNe, likely more diffuse nebulae. 

\begin{figure*}
\includegraphics[width=5.8cm]{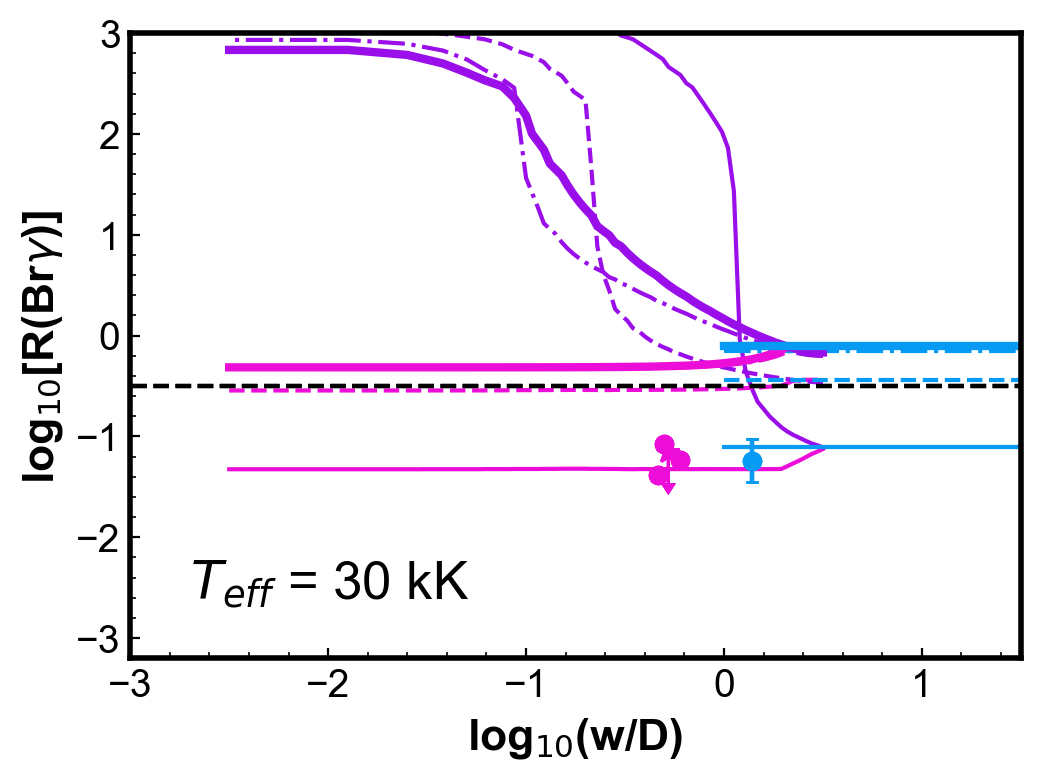}
\includegraphics[width=5.8cm]{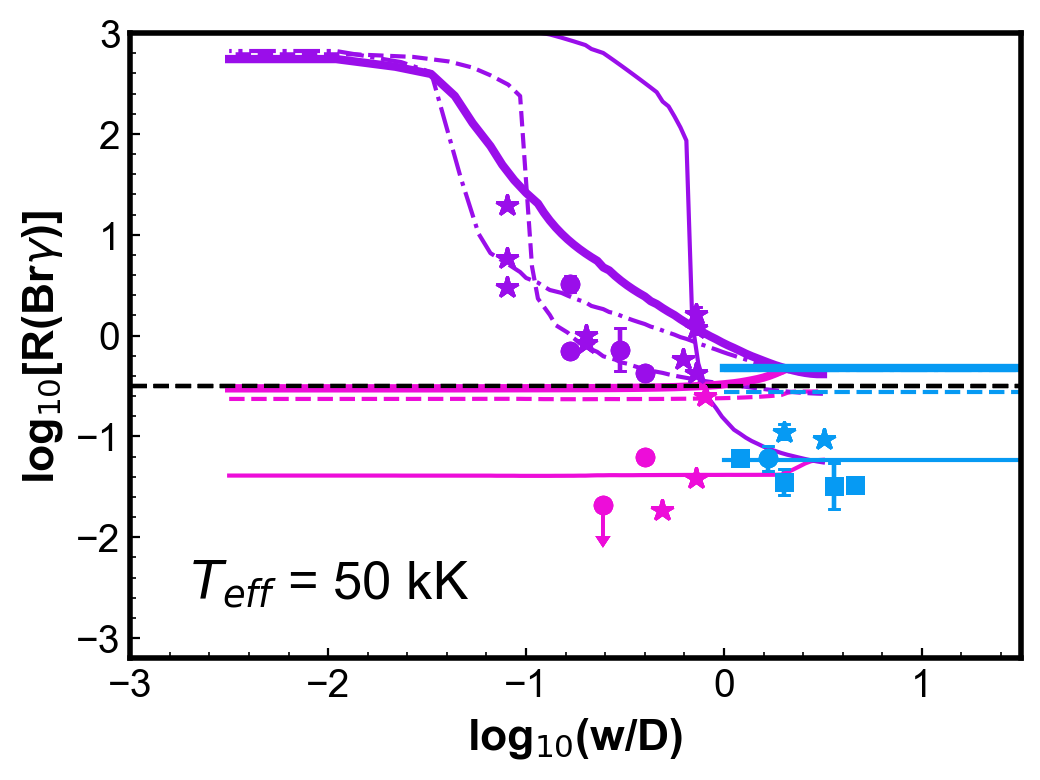}
\includegraphics[width=5.8cm]{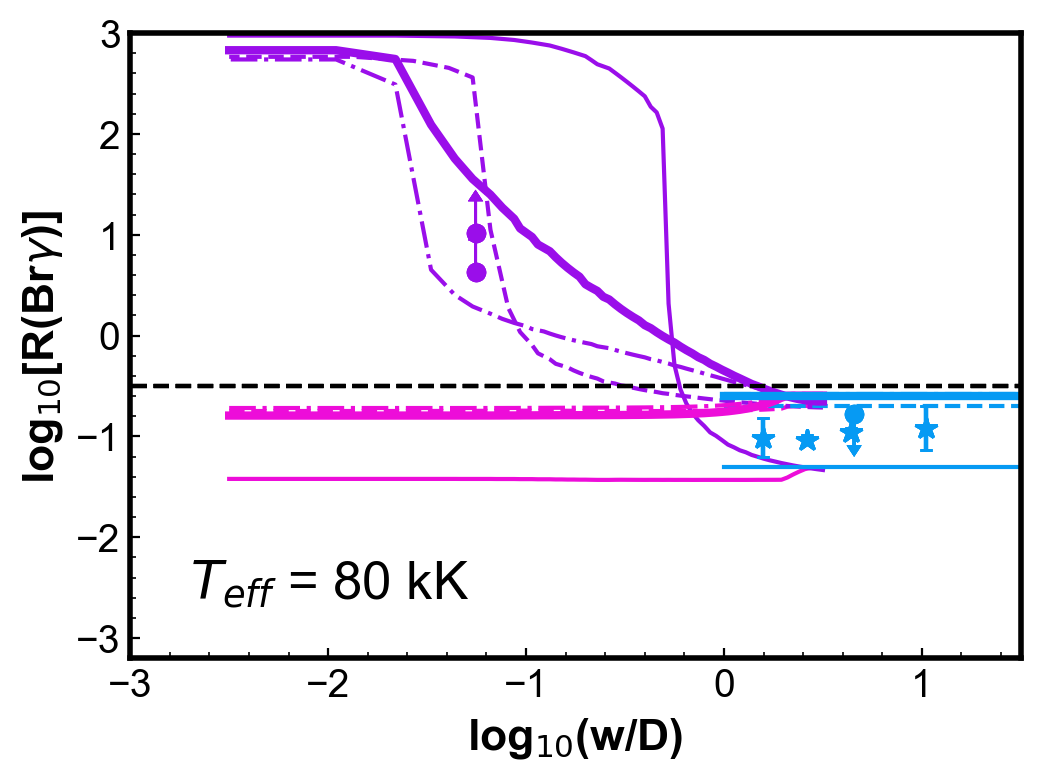}
\includegraphics[width=5.8cm]{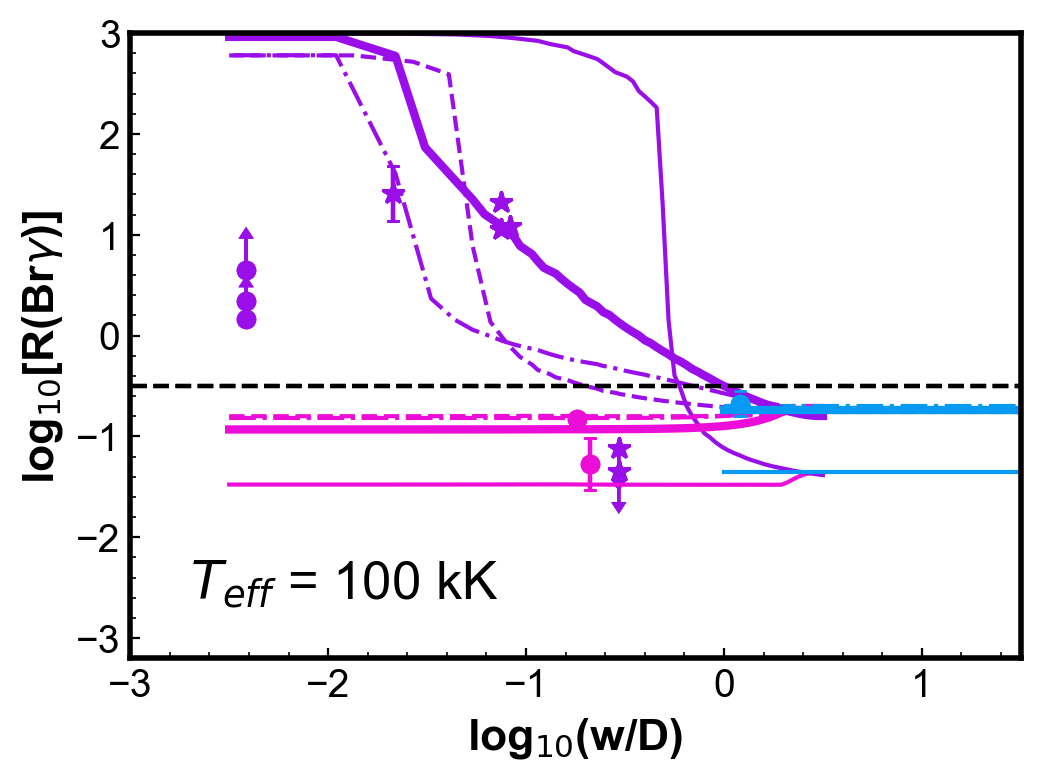}
\includegraphics[width=5.8cm]{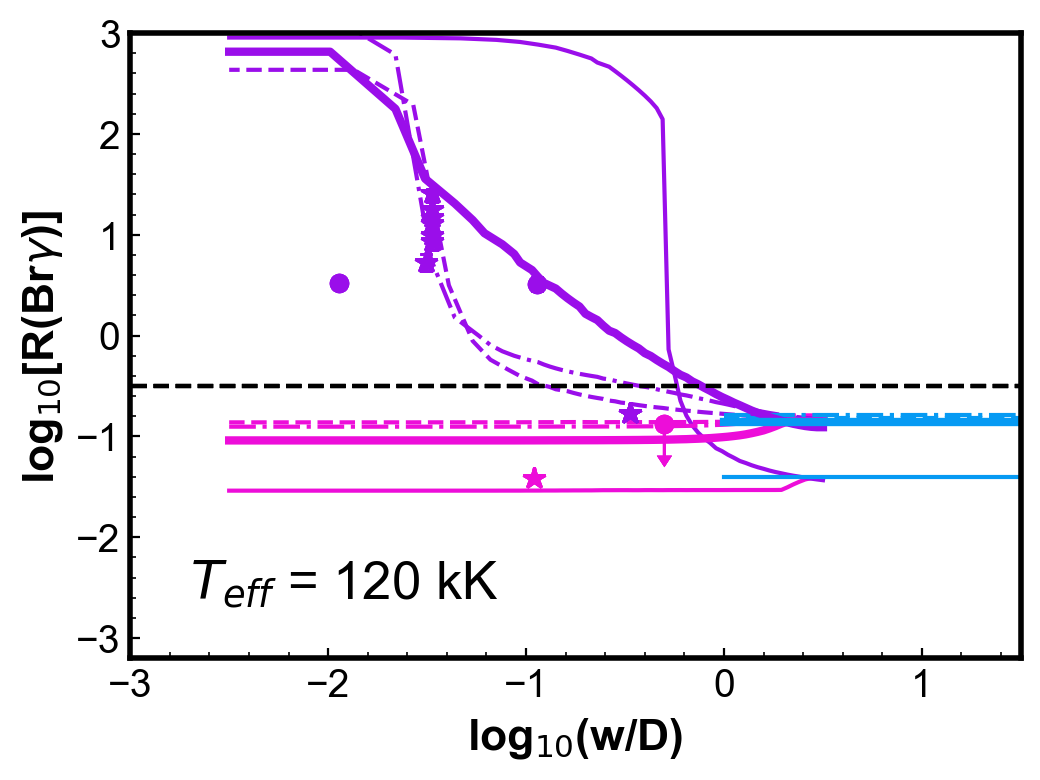}
\includegraphics[width=5.8cm]{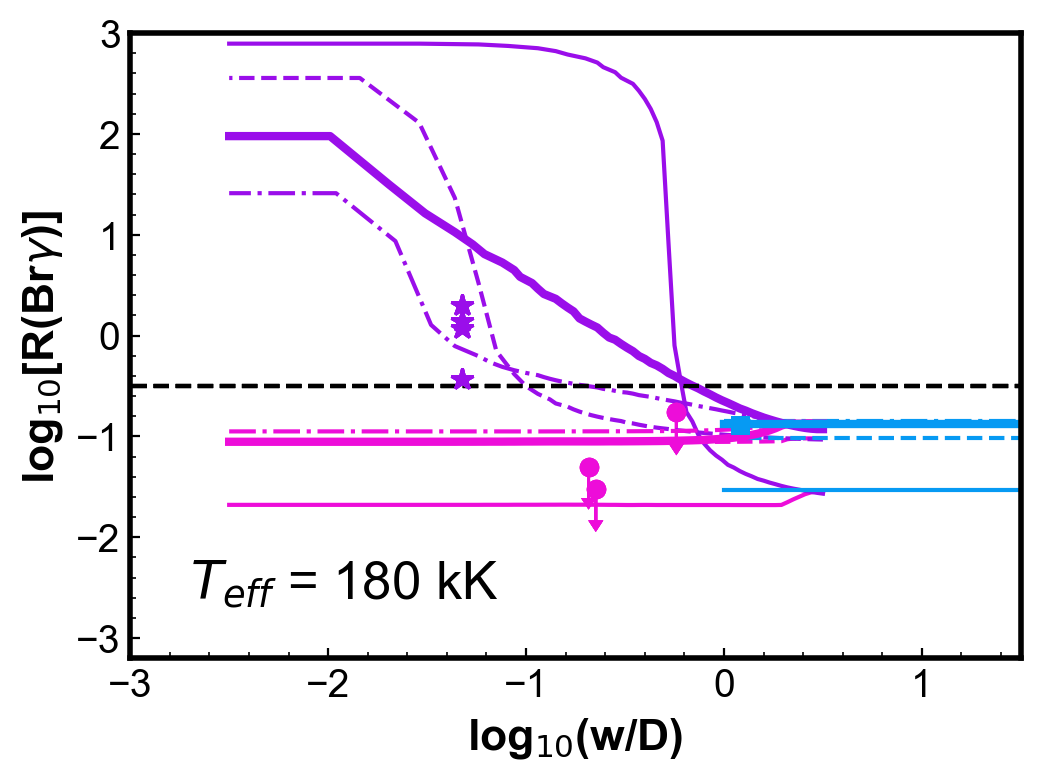}

\caption{Ratio \Rb\ as a function of the ratio $w/D$. Curves in each panel represent shell models with the $T_\textrm{eff}$ indicated and the parameters as in Table~\ref{tab:stparam}. Curves are presented for models with shells simulated at ionization degree of H$^0$/H = 10$^{-4}$ (solid thin), 10$^{-3}$ (dashed), 10$^{-2}$ (dot-dashed), and 5$\times$10$^{-1}$ (solid thick). Dots are observations of PNe with a similar temperature (within 10~kK). The colour code is the same as in Fig~\ref{fig:correlation}. The black dashed line indicates the upper value found for the \textit{whole nebula} configuration.}

\label{fig:ModelsShell1}

\end{figure*}

Molecular hydrogen emission is often associated with dense clumpy shells or torus structures \citep{kastner_etal_1996, matsuura_etal_2007, MarquezLugo_etal_2015, akras_etal_2017}. For this reason, simulations were also made for PNe models where a diffuse gas is surrounded by a dense shell, as describe in Sect.~\ref{sec:models}. In Fig.~\ref{fig:ModelsShell1}, results are shown for shells placed at different distances from the central star. Results are analogous to those from Fig.~\ref{fig:ModelsSphere}. They also reproduce the observations qualitatively and quantitatively. The comparison with observations shows improvement in the match of observations and models with relation to the uniform density models, especially for \textit{centre} and \textit{whole nebula} ratios.

Models with shell internal radius at H$^0$/H = 10$^{-4}$ (solid thin) represents well most \textit{whole nebula} and \textit{centre} observations. On the other hand, models with shell at H$^0$/H = 10$^{-3}$, 10$^{-2}$, and 5$\times$10$^{-1}$, reproduces well most of the \textit{\hh\ peak} observations, as well as some of the other configurations. This may be reflecting the ionization structures being probed by the observations. The \textit{\hh\ peak} configuration probes structures farther form the central star, while \textit{whole nebula} and \textit{centre} is likely to have a stronger influence of the more ionized emission.

The simulations presented here shows that simple photoionization models can reproduce the general behaviour with $w/D$ and the range of observed values if they consider the slit configuration.

\subsection{On \Rb\ as an \hh\ Excitation Mechanism Diagnostic} \label{sec:diagrML}

\begin{figure}

\includegraphics[width=7.9cm]{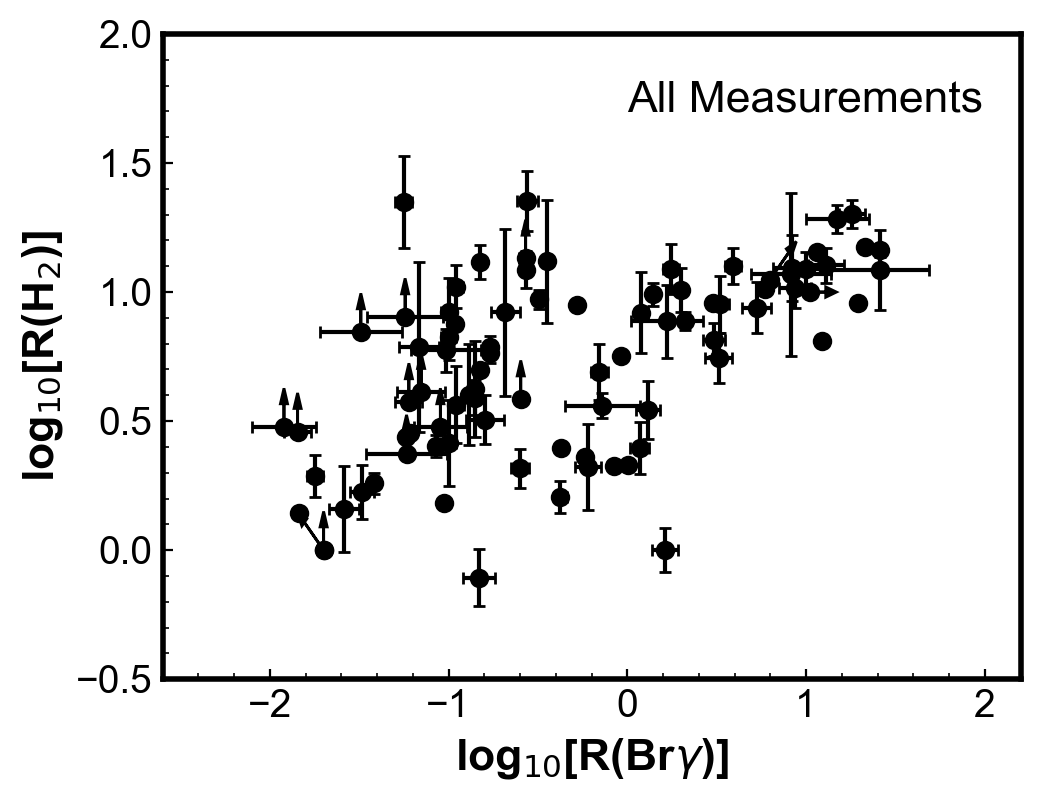}

\includegraphics[width=7.9cm]{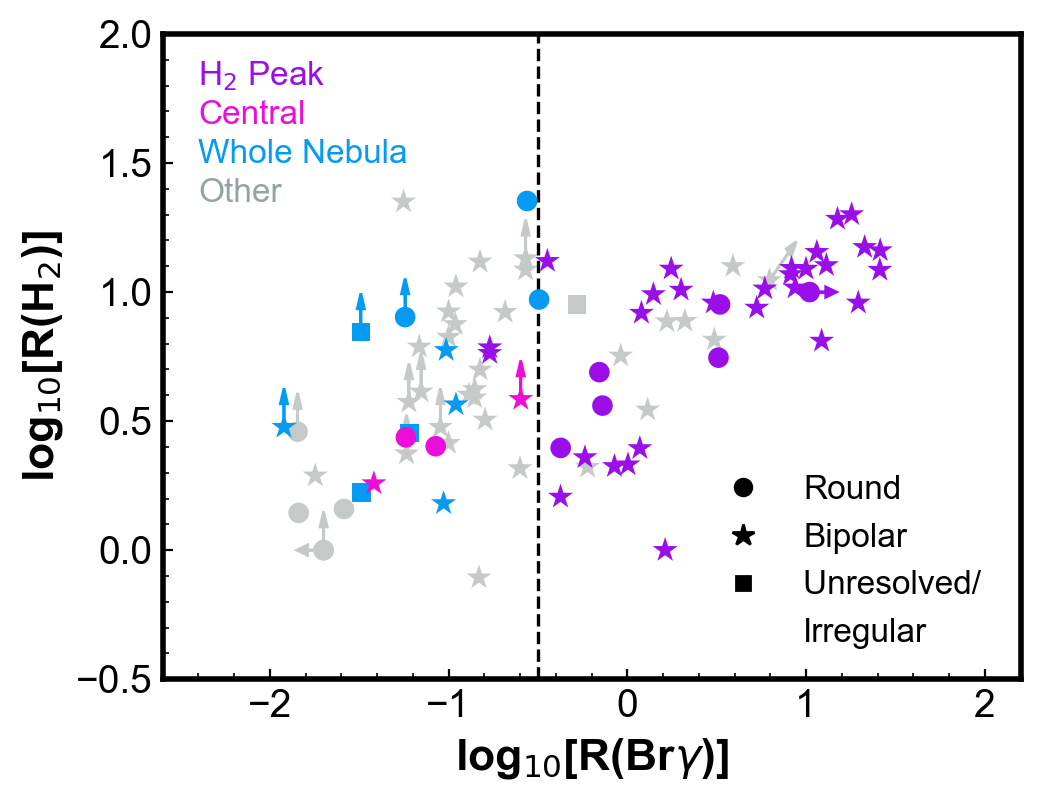}

\includegraphics[width=7.9cm]{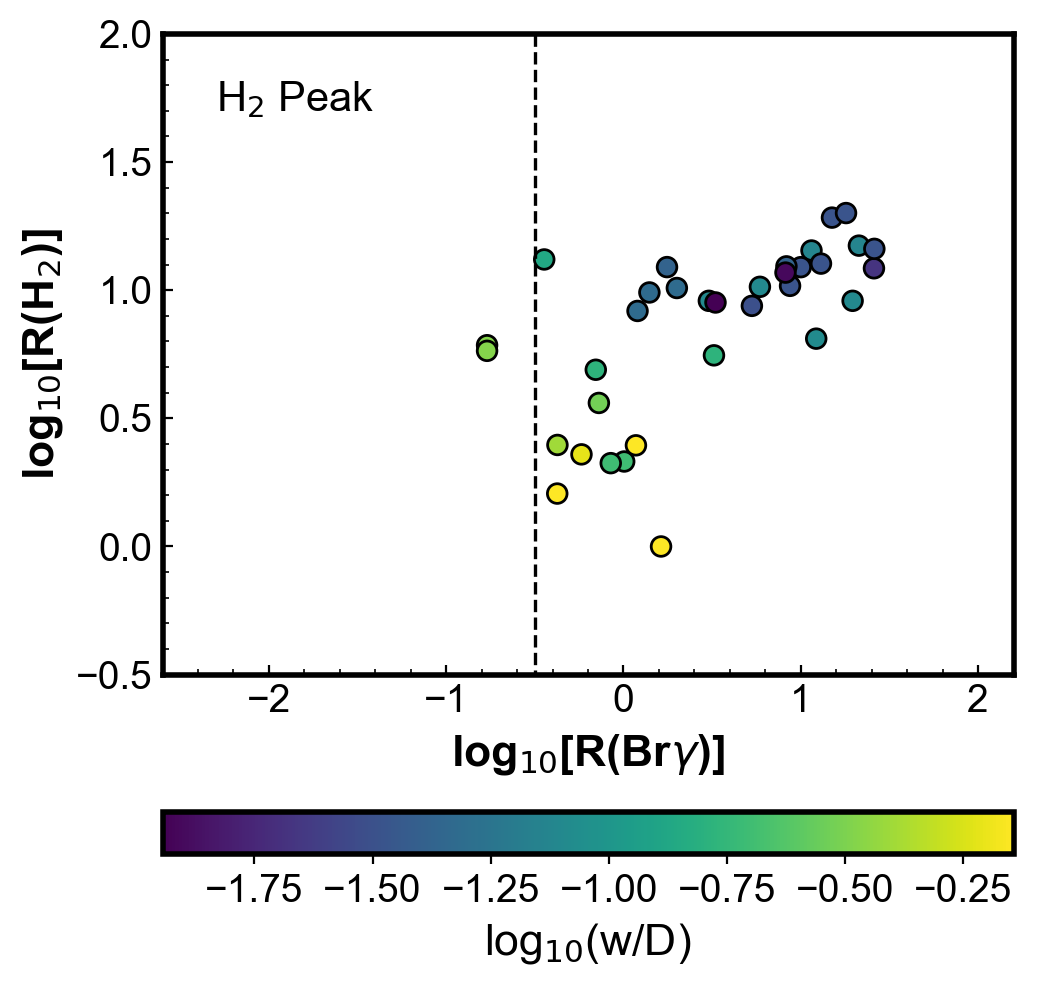}

\caption{Diagnostic diagram proposed by \citet{MarquezLugo_etal_2015}. Observations shown are from Table~\ref{tab:observations}. \textit{Top:} All measurements showing error bars when uncertainties are available. Down/up arrows associated with symbols indicate upper/lower limits.  \textit{Middle:} All measurements classified by slit configuration and object morphology. Error bars are omitted for clarity. \textit{Bottom:} Only the \textit{\hh\ peak} observations are shown. The dot colour code indicates the parameter log$_{10}(w/D)$. Observations with small $w/D$ occupy the region where \citet{MarquezLugo_etal_2015} attribute to shocks. The black dashed line indicates the upper value found for \Rb\ in the \textit{whole nebula} configuration.}

\label{fig:diagramML15}

\end{figure}

\citet{MarquezLugo_etal_2015} proposed the diagram \Rh\ vs. \Rb\ to analyse the excitation mechanism of \hh. Figure~\ref{fig:diagramML15} shows this diagram for the data in Table~\ref{tab:observations}. In the top plot, all observations with both ratios available are shown with error bars when available. The loose positive correlation seen by \citet{MarquezLugo_etal_2015} is also observed here. In the plot, there is indication of two different populations, which become clear in the middle panel, where the observations are separated according to the slit configuration. The segregation in the \Rb\ values for different configurations seen in Fig. ~\ref{fig:correlation} is also seen in the diagram. Observations including the \textit{whole nebula} and with the slit \textit{centred} exhibit only \Rb\ values smaller than 0.3, as previously shown. Values \Rb\ > 0.3 are obtained by \textit{\hh\ peak} observations or \textit{other} configurations. For both \textit{\hh\ peak} and \textit{whole nebula}, a positive correlation between \Rh\ and \Rb\ is seen. No conclusion can be made for \textit{centre} observations, as there are only a few points. All configurations show a similar range of \Rh, which indicates that both \hh\ lines are produced in the same or very similar regions, as shown in Fig~\ref{fig:emiss_stop}.

The bottom plot in Fig.~\ref{fig:diagramML15} shows that the \textit{\hh\ peak} points with both \Rh\ and \Rb\ high values are observed with less nebular area cover by the slit (lower $w/D$). From their diagram, \citet{MarquezLugo_etal_2015} conclude that shock excitation is the dominant mechanism when \Rh\,>~1. They based their conclusion in two arguments: (i) the thermalisation of the \hh\ emission of such objects (when \Rh\,$\sim$\,10) and (ii), in their words, ``that shock-excitation (...) is a much more efficient excitation mechanism than UV fluorescence and thus produces higher levels of emission in the H$_2$ lines''. 
However, according to the results shown here, taking into consideration the slit position in the photoionization simulations such values can easily be reached. This is not an argument against shock excitation, which cannot be discard, but the results presented here show that the observations with higher \Rb\ are biased by the slit effect and argument (ii) above is not valid. If photoprocesses dominate the \hh\ excitation, high density gas could explain the high values \Rh\ usually attributed to shocks. Models by \citet{Sternberg_Dalgarno_1989} showed that UV excitation (not only shocks) may thermalize the \hh\ level population for sufficiently high optical depths \citep[see also discussions in][]{1994ApJ...437..281H, Akras_etal_2020}. \citet{MarquezLugo_etal_2015} suggested to separate UV excitation from shocks do not necessary follows.


\section{Conclusions} \label{sec:conclusions}

This paper presented the analysis of the slit configuration effect on the \Rb\ ratio in PNe, using observations and numerical simulations. The main results are:

\begin{itemize}

\item The \hh\,1-0~S(1) and \brg\ lines are produced in different regions in the nebula, and, therefore, the slit configuration used in the spectroscopic observations strongly affects the \Rb\ ratio. 

\item For round and ring-like PNe, \Rb\ ratios obtained with a slit across the entire nebula passing by its centre (\textit{centred}) provides similar values to those obtained by integrating the line flux over the \textit{whole nebula}. Similar result is obtained for bipolar PNe when observing with the slit across the equatorial region, perpendicular to the main nebular axis (also considered here \textit{centred}).

\item The \Rb\ ratios derived from observations in the \textit{centred} configuration or when the \textit{whole nebula} is integrated reach only values up to 0.3. Values higher than this are only obtained when the slit is positioned at the \textit{\hh\ peak} positions.

\item The \Rb\ ratio measured with the slit at the \textit{\hh\ peak} emission depends on the fraction of the nebula covered by the slit. The largest values of \Rb\ are found when the slit covers a small fraction of the nebula. 

\item When the slit configuration is taken into account, the simple photoionization models presented here can represent very well the range of values and the general behaviour of the observed \Rb. 

\item The result above shows that the argument that shocks are needed to explain the higher values of \Rb\ is not valid. Therefore, this ratio is not a good indicator of the \hh\ excitation mechanism as suggested by \citet{MarquezLugo_etal_2015}. 

\item All the results showed here demonstrate the importance of considering the slit configuration in studies involving the \Rb\ ratio. 

\end{itemize}

It is important to notice that analogous results could be obtained for shock models if they produce similar \hh\ and \brg\ emission distribution. It is not the aim of this work to defend photoprocesses over shocks as the dominant \hh\ excitation mechanism in PNe. Here the focus is on geometrical effects and the intention is just to bring the attention to the effect of the slit configuration.

\section*{Acknowledgements}

This study was financed in part by the Coordena\c{c}\~{a}o de Aperfei\c{c}oamento de Pessoal de N\'{i}vel Superior - Brasil (CAPES) - Finance Code 001. The author is thankful to S. Akras and H. Monteiro for useful discussions and their suggestions to improve this manuscript. This research has made use of the NASA's Astrophysics Data System and the SIMBAD database, operated at CDS, Strasbourg, France \citep{cds_simbad}. This work has also made use of the computing facilities available at the Laboratory of Computational Astrophysics of the Universidade Federal de Itajub\'{a} (LAC-UNIFEI). The LAC-UNIFEI is maintained with grants from CAPES, CNPq and FAPEMIG.

\section*{Data Availability}

The data underlying this article are available in the article. Any additional information can be request to the Author.


\bibliographystyle{mnras}
\bibliography{mainH2}


\appendix

\section{Values and references of the H$_2$ Observations Used in the Text} \label{Appendix:A}

Table \ref{tab:observations} lists the \hh\ observations from the literature used in this paper. Column 1 shows the name of the PN and the code for the position observed according to the authors of the original paper, cited in Column 10. Column 2 and 3 gives the stellar temperature and its reference, respectively. \Rh\ and \Rb, with errors when available, are given in Columns 4 to 7. The slit width and position used in the \hh\ observation are given in Columns 8 and 9, respectively. The PN morphological classification, the value assumed for its diameter, and their references are given in Column 11 to 13.

Most of the stellar temperature values are of Zanstra temperatures, with preference given for values inferred from He~\textsc{II} lines. If effective temperatures are determined from direct star observations, preference is given for such value. If those are not available, then values determined from other methods are used. 

The morphology listed in the table is a general classification based on the published literature cited in Table~\ref{tab:observations} (which is listed below). The objects with ring morphology in which a bipolar structure is not clear are considered here in the class of round nebulae. Whether such nebula is really round or torus-like (thus bipolar) will not affect the result given that the ionization structure is being taken in account during the current analysis.

\begin{table*}
\tiny
\centering
\caption{Observations Database}
\label{tab:observations}
\begin{tabular}{lccccccccccccc}
\hline

Object	&	$T_\star$	&	$T_\star$	&	$R$(H$_2$)	&	$R$(H$_2$)	&	$R$(Br$\gamma$)	&	$R$(Br$\gamma$)	&	$w$	&	Slit	&	H$_2$	&	Morph.	&	$D$	&	$D$	\\
	&	(10$^3$~K)	&	Ref.	&		&	Error	&		&	Error	&	(arcsec)	&	Position	&	Ref.	&		&	(arcsec)	&	Ref.	\\

\hline

BD+30$^o$3639		&	45.7	&	Ph03	&		&		&	0.060	&	0.017	&	10.0	&	Whole	&	Be78	&	R	&	6.0	&	Ha97	\\
BD+30$^o$3639		&	45.7	&	Ph03	&	1.4	&	0.6	&	0.026	&	0.005	&	5.0	&	Other	&	Ge91	&	R	&	6.0	&	Ha97	\\
BD+30$^o$3639	 OE	&	45.7	&	Ph03	&	4.9	&	1.2	&	0.698	&	0.079	&	1.0	&	\hh\ Peak	&	Ho99	&	R	&	6.0	&	Ha97	\\
BD+30$^o$3639	 N Ring	&	45.7	&	Ph03	&		&		&	0.033	&	0.016	&	1.0	&	Other	&	Ho99	&	R	&	6.0	&	Ha97	\\
BD+30$^o$3639	 H$_2$ Lobe	&	45.7	&	Ph03	&	5.6	&	1.2	&	3.238	&	0.569	&	1.0	&	\hh\ Peak	&	Ho99	&	R	&	6.0	&	Ha97	\\
BD+30$^o$3639	 Core	&	45.7	&	Ph03	&	1.4	&		&	0.015	&		&	2.4	&	Other	&	Lu01	&	R	&	6.0	&	Ha97	\\
BD+30$^o$3639	 Nebula	&	45.7	&	Ph03	&		&		&	0.062	&		&	2.4	&	Centre	&	Lu01	&	R	&	6.0	&	Ha97	\\
BD+30$^o$3639	 H$_2$ Zone	&	45.7	&	Ph03	&	2.5	&		&	0.424	&		&	2.4	&	\hh\ Peak	&	Lu01	&	R	&	6.0	&	Ha97	\\
BD+30$^o$3639	 E	&	45.7	&	Ph03	&	3.6	&	0.4	&	0.726	&	0.351	&	1.8	&	\hh\ Peak	&	Li06	&	R	&	6.0	&	Ha97	\\
Cn 3-1		&	53.6	&	Ph03	&		&		&	< 0.015	&		&	1.8	&	Other	&	Li06	&	B	&	3.0	&	Mi97	\\
Hb 5		&	131.0	&	Ph03	&		&		&	< 0.200	&		&	5.0	&	Other	&	We88	&	B	&	18.0	&	Ty03	\\
Hb 5	up	&	131.0	&	Ph03	&	13.1	&	2.0	&	0.1500	&	0.010	&	2.0	&	Other	&	Da03	&	B	&	18.0	&	Ty03	\\
Hb 5	cen	&	131.0	&	Ph03	&	2.6	&	1.0	&	0.100	&	0.010	&	2.0	&	Other	&	Da03	&	B	&	18.0	&	Ty03	\\
Hb 5	dn	&	131.0	&	Ph03	&	12.2	&	2.0	&	0.270	&	0.010	&	2.0	&	Other	&	Da03	&	B	&	18.0	&	Ty03	\\
Hb 12		&	45.5	&	Ph03	&		&		&	0.111	&	0.019	&	10.0	&	Whole	&	Be78	&	B	&	5.0	&	ML15	\\
Hb 12		&	45.5	&	Ph03	&	1.0	&	0.2	&	1.629	&	0.273	&	3.6	&	\hh\ Peak	&	Di88	&	B	&	5.0	&	ML15	\\
Hb 12		&	45.5	&	Ph03	&	2.3	&	0.1	&	0.579	&	0.018	&	3.1	&	\hh\ Peak	&	Ra93	&	B	&	5.0	&	ML15	\\
Hb 12	Core	&	45.5	&	Ph03	&		&		&	< 0.003	&		&	1.0	&	Other	&	HL96	&	B	&	5.0	&	ML15	\\
Hb 12	3.7'' E	&	45.5	&	Ph03	&	2.1	&		&	1.009	&		&	1.0	&	\hh\ Peak	&	HL96	&	B	&	5.0	&	ML15	\\
Hb 12	3.7'' E 2'' S	&	45.5	&	Ph03	&	2.1	&		&	0.847	&		&	1.0	&	\hh\ Peak	&	HL96	&	B	&	5.0	&	ML15	\\
Hb 12	central 1	&	45.5	&	Ph03	&	1.9	&	0.4	&	0.018	&	0.002	&	3.6	&	Other	&	LR96	&	B	&	5.0	&	ML15	\\
Hb 12	central 2	&	45.5	&	Ph03	&	1.8	&	0.2	&	0.038	&	0.002	&	3.6	&	Centre	&	LR96	&	B	&	5.0	&	ML15	\\
Hb 12	west	&	45.5	&	Ph03	&	2.5	&	0.6	&	1.175	&	0.142	&	3.6	&	\hh\ Peak	&	LR96	&	B	&	5.0	&	ML15	\\
Hb 12	east	&	45.5	&	Ph03	&	1.6	&	0.2	&	0.423	&	0.033	&	3.6	&	\hh\ Peak	&	LR96	&	B	&	5.0	&	ML15	\\
Hb 12	PA -5$^o$ Ring	&	45.5	&	Ph03	&	0.8	&	0.2	&	0.148	&	0.030	&	0.75	&	Other	&	ML15	&	B	&	5.0	&	ML15	\\
Hb 12	PA -5$^o$ Envelope	&	45.5	&	Ph03	&	2.1	&	0.4	&	0.251	&	0.028	&	0.75	&	Other	&	ML15	&	B	&	5.0	&	ML15	\\
He 2-111		&	196.9	&	PM89	&		&		&	1.000	&		&	5.0	&	Other	&	We88	&	B	&	74.0	&	Lo93	\\
He 2-114		&	135.0	&	KJ89	&		&		&	> 1.000	&		&	5.0	&	Other	&	We88	&	B	&	25.0	&	We88	\\
He 3-1357		&	45.6	&	Ot17	&	1.5	&		&	0.094	&		&	4.5	&	Whole	&	GH02	&	B	&	1.4	&	GH02	\\
Hf 48	Hen 2-60	&	219.4	&	PM91	&		&		&	> 1.000	&		&	5.0	&	Other	&	We88	&	B	&	21.0	&	We88	\\
Hu 1-2		&	111.0	&	Ph03	&		&		&	0.039	&		&	1.2	&	Centre	&	Lu01	&	B	&	11.0	&	Fa15	\\
IC 418	Centre	&	44.5	&	Ph03	&		&		&	< 0.042	&		&	5.0	&	Other	&	St84	&	E	&	15.0	&	RL12	\\
IC 2003		&	99.8	&	Ph03	&		&		&	< 0.030	&		&	5.0	&	Other	&	Ge91	&	R	&	8.6	&	Fe97	\\
IC 2003	Center	&	99.8	&	Ph03	&		&		&	0.034	&	0.040	&	1.0	&	Other	&	Ho99	&	R	&	8.6	&	Fe97	\\
IC 2003		&	99.8	&	Ph03	&		&		&	0.053	&	0.031	&	1.8	&	Centre	&	Li06	&	R	&	8.6	&	Fe97	\\
IC 2165		&	190.0	&	Ph03	&	> 1.0	&		&	0.020	&		&	5.0	&	Other	&	Ge91	&	R	&	8.0	&	Mi18	\\
IC 2165		&	190.0	&	Ph03	&		&		&	< 0.030	&		&	1.8	&	Centre	&	Li06	&	R	&	8.0	&	Mi18	\\
IC 4406	Centre	&	96.8	&	Ph03	&		&		&	1.351	&		&	5.0	&	Other	&	St84	&	B	&	30.0	&	St84	\\
IC 4997		&	62.0	&	Ph03	&		&		&	< 0.200	&		&	10.0	&	Whole	&	Be78	&	B	&	2.2	&	ML15	\\
IC 4997		&	62.0	&	Ph03	&	> 3.0	&		&	0.012	&	0.005	&	5.0	&	Whole	&	Ge91	&	B	&	2.2	&	ML15	\\
IC 5117		&	82.6	&	Ph03	&		&		&	0.120	&	0.060	&	12.0	&	Whole	&	Is84	&	M	&	1.1	&	Hs14	\\
IC 5117		&	82.6	&	Ph03	&	3.7	&	1.3	&	0.110	&	0.011	&	5.0	&	Whole	&	Ge91	&	M	&	1.1	&	Hs14	\\
IC 5117		&	82.6	&	Ph03	&		&		&	0.093	&	0.009	&	3.0	&	Whole	&	Ru01	&	M	&	1.1	&	Hs14	\\
IC 5117		&	82.6	&	Ph03	&	6.0	&	1.1	&	0.097	&	0.044	&	1.8	&	Whole	&	Li06	&	M	&	1.1	&	Hs14	\\
IC 5217		&	78.2	&	Ph03	&		&		&	< 0.008	&		&	1.8	&	Other	&	Li06	&	B	&	6.6	&	Hy01	\\
J900		&	106.0	&	Ph03	&		&		&	0.210	&	0.060	&	12.0	&	Whole	&	Is84	&	R	&	10.0	&	Sh95	\\
J900		&	106.0	&	Ph03	&		&		&	0.148	&	0.022	&	1.8	&	Centre	&	Li06	&	R	&	10.0	&	Sh95	\\
K 3-60		&	185.0	&	Lu01	&		&		&	0.128	&		&	1.2	&	Whole	&	Lu01	&	I	&	1.0	&	St16	\\
K 3-67		&	59.2	&	Lu01	&		&		&	0.019	&		&	1.2	&	Centre	&	Lu01	&	B	&	2.5	&	Sa11	\\
K 4-48		&	125.0	&	Lu01	&	8.9	&		&	0.523	&		&	1.2	&	Other	&	Lu01	&	?	&	2.2	&	St08	\\
LMC SMP-06		&	140	&	Ma16	&		&		&	0.098	&	0.006	&	0.75	&	Whole	&	Ma16	&	E	&	0.67	&	Sh06	\\
LMC SMP-47		&	150	&	Ma16	&	9.4	&	0.8	&	0.321	&	0.009	&	0.75	&	Whole	&	Ma16	&	E	&	0.45	&	Sh06	\\
LMC SMP-62		&	100	&	Ma16	&	> 2.9	&		&	0.014	&	0.002	&	0.45	&	Other	&	Ma16	&	E	&	0.59	&	Sh06	\\
LMC SMP-73		&	135	&	Ma16	&	22.5	&	6.1	&	0.275	&	0.037	&	0.75	&	Whole	&	Ma16	&	E	&	0.31	&	Sh06	\\
LMC SMP-85		&	46	&	Ma16	&	1.7	&	0.4	&	0.033	&	0.005	&	0.75	&	Whole	&	Ma16	&	U	&	<0.163	&	HB04	\\
LMC SMP-99		&	124	&	Ma16	&		&		&	0.037	&	0.010	&	0.75	&	Other	&	Ma16	&	B	&	0.82	&	St07	\\
M 1-4		&	40.3	&	Ph03	&		&		&	< 0.02	&		&	1.8	&	Centre	&	Li06	&	R	&	7.4	&	Sa11	\\
M 1-6		&	34.5	&	Ph03	&		&		&	< 0.07	&		&	1.8	&	Centre	&	Li06	&	B	&	3.4	&	Sa11	\\
M 1-11		&	32.0	&	Ot13	&	2.7	&		&	0.058	&		&	1.2	&	Centre	&	Lu01	&	R	&	2.0	&	Ot13	\\
M 1-11		&	32.0	&	Ot13	&	2.5	&	0.2	&	0.085	&	0.006	&	1.0	&	Centre	&	Ot13	&	R	&	2.0	&	Ot13	\\
M 1-13		&	118.1	&	PM91	&	7.7	&	2.5	&	1.667	&	0.767	&	1.8	&	Other	&	Li06	&	B	&	12.0	&	Li06	\\
M 1-74		&	58.0	&	Ph03	&	2.8	&		&	0.061	&		&	1.2	&	Whole	&	Lu01	&	U	&	1.0	&	SS99	\\

\hline
\end{tabular}
\end{table*}

\setcounter{table}{0}

\begin{table*}
\centering
\tiny
\caption{(Cont) Observations Database}
\label{tab:observations2}
\begin{tabular}{lccccccccccccc}
\hline

Object	&	T$\star$	&	T$\star$	&	$R$(H$_2$)	&	$R$(H$_2$)	&	$R$(Br$\gamma$)	&	$R$(Br$\gamma$)	&	$w$	&	Slit	&	H$_2$	&	Morph.	&	$D$	&	$D$	\\
	&	(10$^3$~K)	&	Ref.	&		&	Error	&		&	Error	&	(arcsec)	&	Position	&	Ref.	&		&	(arcsec)	&	Ref.	\\

\hline

M 1-75	Lobes	&	200.0	&	Hu88	&	12.4	&	3.7	&	8.300	&	2.033	&	0.75	&	\hh\ Peak	&	ML15	&	M	&	18.0	&	Gu00	\\
M 1-75	Ring	&	200.0	&	Hu88	&	12.3	&	2.8	&	1.760	&	0.179	&	0.75	&	\hh\ Peak	&	ML15	&	M	&	18.0	&	Gu00	\\
M 1-75	Centre	&	200.0	&	Hu88	&		&		&	0.477	&	0.037	&	0.75	&	Other	&	ML15	&	M	&	18.0	&	Gu00	\\
M 2-9		&	43.3	&	Ph03	&	> 3.9	&		&	0.254	&		&	8.0	&	Centre	&	Ph85	&	B	&	10.0	&	HL94	\\
M 2-9	Core	&	43.3	&	Ph03	&		&		&	< 0.001	&		&	0.8	&	Other	&	HL94	&	B	&	10.0	&	HL94	\\
M 2-9	N knot	&	43.3	&	Ph03	&	5.0	&		&	0.150	&		&	0.8	&	Other	&	HL94	&	B	&	10.0	&	HL94	\\
M 2-9	Lobe I	&	43.3	&	Ph03	&	9.1	&		&	3.030	&		&	0.8	&	\hh\ Peak	&	HL94	&	B	&	10.0	&	HL94	\\
M 2-9	Lobe M	&	43.3	&	Ph03	&	10.3	&		&	5.882	&		&	0.8	&	\hh\ Peak	&	HL94	&	B	&	10.0	&	HL94	\\
M 2-9	Lobe O	&	43.3	&	Ph03	&	9.1	&		&	19.61	&		&	0.8	&	\hh\ Peak	&	HL94	&	B	&	10.0	&	HL94	\\
M 4-17	PA 130$^o$ All	&	127.0	&	St02	&	6.5	&	1.0	&	3.070	&	0.436	&	0.75	&	Other	&	ML15	&	B	&	24.0	&	Gu00	\\
M 4-17	PA 40$^o$ Ring	&	127.0	&	St02	&	8.7	&	2.0	&	5.300	&	0.986	&	0.75	&	\hh\ Peak	&	ML15	&	B	&	24.0	&	Gu00	\\
M 4-17	PA 40$^o$ Centre	&	127.0	&	St02	&	7.7	&	0.6	&	2.100	&	0.123	&	0.75	&	Other	&	ML15	&	B	&	24.0	&	Gu00	\\
Me 2-1		&	180.0	&	Ph03	&		&		&	< 0.170	&		&	5.0	&	Centre	&	St84	&	R	&	8.7	&	Su04	\\
Me 2-1		&	180.0	&	Ph03	&		&		&	< 0.050	&		&	1.8	&	Centre	&	Li06	&	R	&	8.7	&	Su04	\\
MyCn 18		&	51.6	&	Ph03	&		&		&	< 0.200	&		&	5.0	&	Other	&	We88	&	B	&	6.0	&	Cl14	\\
Mz 1		&	139.0	&	KJ89	&		&		&	> 1.000	&		&	5.0	&	Other	&	We88	&	B	&	58.4	&	Ma98	\\
NGC 40	W Lobe	&	33.8	&	Ph03	&	6.1	&	4.7	&	0.069	&	0.018	&	1.0	&	Other	&	Ho99	&	B	&	45.0	&	LF11	\\
NGC 40	W	&	33.8	&	Ph03	&	> 2.4	&		&	0.058	&	0.030	&	1.8	&	Other	&	Li06	&	B	&	45.0	&	LF11	\\
NGC 40	E	&	33.8	&	Ph03	&		&		&	< 0.040	&		&	1.8	&	Other	&	Li06	&	B	&	45.0	&	LF11	\\
NGC 1535	Centre	&	76.3	&	Ph03	&		&		&	< 0.450	&		&	5.0	&	Other	&	St84	&	R	&	35.0	&	BA91	\\
NGC 2346		&	100.0	&	Ma15	&		&		&	5.000	&		&	5.0	&	Other	&	We88	&	B	&	47.0	&	Vi99	\\
NGC 2346	W filament	&	100.0	&	Ma15	&	12.2	&	4.4	&	25.875	&	16.253	&	1.0	&	\hh\ Peak	&	Ho99	&	B	&	47.0	&	Vi99	\\
NGC 2346	W	&	100.0	&	Ma15	&	14.3	&		&	11.494	&		&	3.5	&	\hh\ Peak	&	Vi99	&	B	&	47.0	&	Vi99	\\
NGC 2346	E	&	100.0	&	Ma15	&	14.9	&		&	21.277	&		&	3.5	&	\hh\ Peak	&	Vi99	&	B	&	47.0	&	Vi99	\\
NGC 2346	S	&	100.0	&	Ma15	&	> 11.1	&		&	> 6.250	&		&	3.5	&	Other	&	Vi99	&	B	&	47.0	&	Vi99	\\
NGC 2440		&	200.0	&	HH90	&		&		&	0.160	&	0.040	&	12.0	&	Other	&	Is84	&	B	&	80.0	&	CP00	\\
NGC 2440		&	200.0	&	HH90	&	> 13.5	&		&	0.270	&	0.022	&	5.0	&	Other	&	Ge91	&	B	&	80.0	&	CP00	\\
NGC 2440	NE Clump	&	200.0	&	HH90	&	11.7	&	8.5	&	8.200	&	4.220	&	1.0	&	\hh\ Peak	&	Ho99	&	B	&	80.0	&	CP00	\\
NGC 2440	N Lobe	&	200.0	&	HH90	&	8.4	&	6.2	&	0.207	&	0.039	&	1.0	&	Other	&	Ho99	&	B	&	80.0	&	CP00	\\
NGC 2440	E Lobe	&	200.0	&	HH90	&		&		&	0.627	&	0.205	&	1.0	&	Other	&	Ho99	&	B	&	80.0	&	CP00	\\
NGC 2792	Centre	&	114.2	&	Ph03	&		&		&	< 0.130	&		&	5.0	&	Centre	&	St84	&	R	&	10.0	&	Po09	\\
NGC 2818	20'' S	&	215.0	&	Ph03	&		&		&	> 3.420	&		&	5.0	&	Other	&	St84	&	B	&	57.1	&	Va12	\\
NGC 2818	30'' E	&	215.0	&	Ph03	&		&		&	> 3.540	&		&	5.0	&	Other	&	St84	&	B	&	57.1	&	Va12	\\
NGC 2818		&	215.0	&	Ph03	&		&		&	35.000	&		&	5.0	&	Other	&	We88	&	B	&	57.1	&	Va12	\\
NGC 2899		&	270.0	&	Fr08	&		&		&	3.000	&		&	5.0	&	Other	&	We88	&	B	&	60.0	&	Lo01	\\
NGC 3132	Centre	&	80.1	&	Ph03	&		&		&	> 3.820	&		&	5.0	&	Other	&	St84	&	R	&	90.0	&	Mo00	\\
NGC 3132	20'' N	&	80.1	&	Ph03	&		&		&	> 4.290	&		&	5.0	&	\hh\ Peak	&	St84	&	R	&	90.0	&	Mo00	\\
NGC 3132	20'' E	&	80.1	&	Ph03	&	10.0	&		&	> 10.480	&		&	5.0	&	\hh\ Peak	&	St84	&	R	&	90.0	&	Mo00	\\
NGC 3132		&	80.1	&	Ph03	&		&		&	10.50	&		&	5.0	&	Other	&	We88	&	R	&	90.0	&	Mo00	\\
NGC 3242	Centre	&	89.9	&	Ph03	&		&		&	< 0.080	&		&	5.0	&	Other	&	St84	&	R	&	20.0	&	BK18	\\
NGC 3242	15'' E	&	89.9	&	Ph03	&		&		&	< 0.170	&		&	5.0	&	Other	&	St84	&	R	&	20.0	&	BK18	\\
NGC 3242		&	89.9	&	Ph03	&		&		&	< 0.007	&		&	5.0	&	Other	&	Ge91	&	R	&	20.0	&	BK18	\\
NGC 4071		&	118.0	&	Ph03	&		&		&	> 1.000	&		&	5.0	&	Other	&	We88	&	B	&	66.7	&	Be17	\\
NGC 5189		&	109.8	&	Ph03	&		&		&	> 1.000	&		&	5.0	&	Other	&	We88	&	P	&	210.0	&	Be17	\\
NGC 6072	Centre	&	147.0	&	Ph03	&		&		&	> 4.210	&		&	5.0	&	Other	&	St84	&	M	&	67.0	&	Kw10	\\
NGC 6072	20''E	&	147.0	&	Ph03	&		&		&	> 5.710	&		&	5.0	&	Other	&	St84	&	M	&	67.0	&	Kw10	\\
NGC 6072	20''W	&	147.0	&	Ph03	&		&		&	> 6.180	&		&	5.0	&	Other	&	St84	&	M	&	67.0	&	Kw10	\\
NGC 6153	Centre	&	97.1	&	Ph03	&		&		&	< 0.110	&		&	5.0	&	Other	&	St84	&	B	&	17.0	&	Yu11	\\
NGC 6153	10'' E	&	97.1	&	Ph03	&		&		&	< 0.050	&		&	5.0	&	\hh\ Peak	&	St84	&	B	&	17.0	&	Yu11	\\
NGC 6153	10'' W	&	97.1	&	Ph03	&		&		&	< 0.080	&		&	5.0	&	\hh\ Peak	&	St84	&	B	&	17.0	&	Yu11	\\
NGC 6210		&	61.1	&	Ph03	&		&		&	0.050	&	0.020	&	5.0	&	Other	&	Is84	&	I	&	15.0	&	Po09b	\\
NGC 6210		&	61.1	&	Ph03	&		&		&	< 0.010	&		&	5.0	&	Other	&	Ge91	&	I	&	15.0	&	Po09b	\\
NGC 6210		&	61.1	&	Ph03	&		&		&	< 0.007	&		&	1.8	&	Other	&	Li06	&	I	&	15.0	&	Po09b	\\
NGC 6302		&	200.0	&	Sz09	&		&		&	0.084	&		&	7.5	&	Other	&	Ph83	&	B	&	97.0	&	Sz11	\\
NGC 6302		&	200.0	&	Sz09	&	> 3.0	&		&	0.090	&	0.030	&	5.0	&	Other	&	Ge91	&	B	&	97.0	&	Sz11	\\
NGC 6302	up2	&	200.0	&	Sz09	&	3.5	&	0.9	&	1.300	&	0.200	&	2.0	&	Other	&	Da03	&	B	&	97.0	&	Sz11	\\
NGC 6302	up	&	200.0	&	Sz09	&	3.2	&	0.7	&	0.160	&	0.040	&	2.0	&	Other	&	Da03	&	B	&	97.0	&	Sz11	\\
NGC 6302	cen	&	200.0	&	Sz09	&	6.7	&	1.4	&	0.100	&	0.010	&	2.0	&	Other	&	Da03	&	B	&	97.0	&	Sz11	\\
NGC 6302	dn	&	200.0	&	Sz09	&	3.9	&	0.6	&	0.140	&	0.010	&	2.0	&	Other	&	Da03	&	B	&	97.0	&	Sz11	\\
NGC 6302	dn2	&	200.0	&	Sz09	&	2.1	&	0.8	&	0.600	&	0.100	&	2.0	&	Other	&	Da03	&	B	&	97.0	&	Sz11	\\
NGC 6302	dn3	&	200.0	&	Sz09	&	12.6	&	2.0	&	3.900	&	0.400	&	2.0	&	Other	&	Da03	&	B	&	97.0	&	Sz11	\\

\hline
\end{tabular}
\end{table*}

\setcounter{table}{0}

\begin{table*}
\centering
\tiny
\caption{(Cont) Observations Database}
\label{tab:observations3}
\begin{tabular}{lccccccccccccc}
\hline

Object		&	T$\star$	&	T$\star$	&	$R$(H$_2$)	&	$R$(H$_2$)	&	$R$(Br$\gamma$)	&	$R$(Br$\gamma$)	&	$w$	&	Slit	&	H$_2$	&	Morph.	&	$D$	&	$D$	\\
	&	(10$^3$~K)	&	Ref.	&		&	Error	&		&	Error	&	(arcsec)	&	Position	&	Ref.	&		&	(arcsec)	&	Ref.	\\

\hline

NGC 6445	up	&	182.7	&	Ph03	&	10.2	&	2.0	&	2.000	&	0.100	&	2.0	&	\hh\ Peak	&	Da03	&	B	&	42.0	&	Da03	\\
NGC 6445	cen	&	182.7	&	Ph03	&	8.3	&	3.0	&	1.200	&	0.100	&	2.0	&	\hh\ Peak	&	Da03	&	B	&	42.0	&	Da03	\\
NGC 6445	dn	&	182.7	&	Ph03	&		&		&	0.370	&	0.020	&	2.0	&	\hh\ Peak	&	Da03	&	B	&	42.0	&	Da03	\\
NGC 6445	dn2	&	182.7	&	Ph03	&	9.8	&	1.0	&	1.400	&	0.100	&	2.0	&	\hh\ Peak	&	Da03	&	B	&	42.0	&	Da03	\\
NGC 6537		&	250.0	&	KJ89	&	> 3.8	&		&	0.060	&	0.010	&	5.0	&	Other	&	Ge91	&	B	&	50.0	&	Da03	\\
NGC 6537	up	&	250.0	&	KJ89	&	4.2	&	1.8	&	0.140	&	0.010	&	2.0	&	Other	&	Da03	&	B	&	50.0	&	Da03	\\
NGC 6537	cen	&	250.0	&	KJ89	&	8.4	&	2.5	&	0.100	&	0.010	&	2.0	&	Other	&	Da03	&	B	&	50.0	&	Da03	\\
NGC 6537	dn	&	250.0	&	KJ89	&	4.0	&	1.8	&	0.130	&	0.010	&	2.0	&	Other	&	Da03	&	B	&	50.0	&	Da03	\\
NGC 6572		&	66.8	&	Ph03	&		&		&	< 0.025	&		&	10.0	&	Centre	&	Be78	&	B	&	7.0	&	Mi99	\\
NGC 6572		&	66.8	&	Ph03	&		&		&	< 0.004	&		&	5.0	&	Other	&	Ge91	&	B	&	7.0	&	Mi99	\\
NGC 6720		&	120.6	&	Ph03	&		&		&	3.238	&	0.500	&	10.0	&	\hh\ Peak	&	Be78	&	R	&	88.0	&	OD03	\\
NGC 6720	L	&	120.6	&	Ph03	&	9.0	&	2.3	&	3.303	&	0.362	&	1.0	&	\hh\ Peak	&	Ho99	&	R	&	88.0	&	OD03	\\
NGC 6772		&	119.9	&	Ph03	&		&		&	> 1.000	&		&	5.0	&	Other	&	We88	&	R	&	90.0	&	Fa18	\\
NGC 6778		&	96.9	&	Ph03	&		&		&	< 1.000	&		&	5.0	&	Other	&	We88	&	B	&	20.0	&	GM12	\\
NGC 6790		&	74.0	&	Ph03	&		&		&	< 0.170	&		&	10.0	&	Whole	&	Be78	&	R	&	1.8	&	ZK91	\\
NGC 6881	PA 137$^o$ 	&	95.6	&	Ph03	&	7.5	&		&	0.109	&		&	0.75	&	Other	&	RL08	&	B	&	9.0	&	RL08	\\
Central\\
NGC 6881	PA 137$^o$ 	&	95.6	&	Ph03	&		&		&	4.308	&		&	0.75	&	Other	&	RL08	&	B	&	9.0	&	RL08	\\
H$_2$ Lobes\\
NGC 6881	PA 137$^o$ 	&	95.6	&	Ph03	&		&		&	0.855	&		&	0.75	&	Other	&	RL08	&	B	&	9.0	&	RL08	\\
Ion. Lobes\\
NGC 6881	PA 113$^o$ 	&	95.6	&	Ph03	&	6.5	&		&	12.222	&		&	0.75	&	\hh\ Peak	&	RL08	&	B	&	9.0	&	RL08	\\
H$_2$ Lobes\\
NGC 6881	PA 113$^o$	&	95.6	&	Ph03	&	5.7	&		&	0.919	&		&	0.75	&	Other	&	RL08	&	B	&	9.0	&	RL08	\\
 Ion. Lobes\\
NGC 6886	up	&	129.0	&	Ph03	&	6.1	&	0.6	&	0.170	&	0.010	&	2.0	&	\hh\ Peak	&	Da03	&	B	&	6.0	&	Da03	\\
NGC 6886	cen	&	129.0	&	Ph03	&	10.5	&	2.0	&	0.110	&	0.010	&	2.0	&	Other	&	Da03	&	B	&	6.0	&	Da03	\\
NGC 6886	dn	&	129.0	&	Ph03	&	5.8	&	0.5	&	0.170	&	0.010	&	2.0	&	\hh\ Peak	&	Da03	&	B	&	6.0	&	Da03	\\
NGC 7009		&	87.8	&	Ph03	&		&		&	< 1.000	&		&	5.0	&	Other	&	We88	&	B	&	50.0	&	Wa18	\\
NGC 7027		&	198.0	&	La00	&	> 4.1	&		&	0.070	&	0.022	&	7.0	&	Other	&	Sm81	&	M	&	7.3	&	La16	\\
NGC 7027		&	198.0	&	La00	&	22.4	&	9.2	&	0.056	&	0.006	&	5.0	&	Other	&	Ge91	&	M	&	7.3	&	La16	\\
NGC 7027	NW 	&	198.0	&	La00	&	13.2	&	7.2	&	0.357	&	0.025	&	1.0	&	\hh\ Peak	&	Ho99	&	M	&	7.3	&	La16	\\
H$_2$ Lobe\\
NGC 7027	W Lobe	&	198.0	&	La00	&		&		&	0.051	&	0.014	&	1.0	&	Other	&	Ho99	&	M	&	7.3	&	La16	\\
NGC 7027		&	198.0	&	La00	&		&		&	0.096	&		&	1.2	&	Other	&	Lu01	&	M	&	7.3	&	La16	\\
NGC 7048	up2	&	119.5	&	Ph03	&	19.2	&	2.4	&	15.000	&	6.000	&	2.0	&	\hh\ Peak	&	Da03	&	B	&	60.0	&	Da03	\\
NGC 7048	up	&	119.5	&	Ph03	&	12.7	&	2.0	&	13.000	&	3.000	&	2.0	&	\hh\ Peak	&	Da03	&	B	&	60.0	&	Da03	\\
NGC 7048	cen	&	119.5	&	Ph03	&	12.3	&	1.8	&	10.000	&	1.800	&	2.0	&	\hh\ Peak	&	Da03	&	B	&	60.0	&	Da03	\\
NGC 7048	dn	&	119.5	&	Ph03	&	20.0	&	2.5	&	18.000	&	3.000	&	2.0	&	\hh\ Peak	&	Da03	&	B	&	60.0	&	Da03	\\
NGC 7048	dn2	&	119.5	&	Ph03	&	10.4	&	1.9	&	8.700	&	1.800	&	2.0	&	\hh\ Peak	&	Da03	&	B	&	60.0	&	Da03	\\
NGC 7048	dn3	&	119.5	&	Ph03	&	14.5	&	0.8	&	26.000	&	2.000	&	2.0	&	\hh\ Peak	&	Da03	&	B	&	60.0	&	Da03	\\
NGC 7293	5 'N	&	108.5	&	Ph03	&		&		&	> 2.190	&		&	5.0	&	\hh\ Peak	&	St84	&	R	&	1300.0	&	He99	\\
NGC 7293	7 'E	&	108.5	&	Ph03	&		&		&	> 1.470	&		&	5.0	&	\hh\ Peak	&	St84	&	R	&	1300.0	&	He99	\\
NGC 7293	7 'W	&	108.5	&	Ph03	&		&		&	> 4.470	&		&	5.0	&	\hh\ Peak	&	St84	&	R	&	1300.0	&	He99	\\
NGC 7662		&	109.9	&	Ph03	&		&		&	0.130	&		&	5.0	&	Other	&	Is84	&	R	&	35.0	&	BK18	\\
NGC 7662		&	109.9	&	Ph03	&	> 1.0	&		&	< 0.020	&		&	5.0	&	Other	&	Ge91	&	R	&	35.0	&	BK18	\\
SwSt 1		&	35.5	&	Ph03	&		&		&	0.042	&		&	0.6	&	Centre	&	DM01	&	R	&	1.3	&	DM01	\\
SwSt 1		&	35.5	&	Ph03	&	> 8.0	&		&	0.057	&	0.028	&	1.8	&	Whole	&	Li06	&	R	&	1.3	&	DM01	\\
Vy 1-2		&	99.8	&	Ph03	&		&		&	< 0.030	&		&	1.8	&	Other	&	Li06	&	B	&	3.0	&	Ak15	\\
Vy 2-2	Core	&	59.5	&	Ph03	&		&		&	0.035	&	0.010	&	1.0	&	Whole	&	Ho99	&	U	&	0.5	&	CS98	\\
Vy 2-2		&	59.5	&	Ph03	&	> 7.0	&		&	0.032	&	0.017	&	1.8	&	Whole	&	Li06	&	U	&	0.5	&	CS98	\\

\hline
\end{tabular}
\end{table*}


\vspace{0.5cm}

\noindent \textbf{References for the \hh\ observations:} 
Be78: \citet{1978ApJ...219L..33B}, 
Da03: \citet{davis_etal_2003}, 
Di88: \citet{1988ApJ...327L..27D}, 
DM01: \citet{2001MNRAS.328..527D}, 
Ge91: \citet{Geballe_etal_1991}, 
GH02: \citet{2002A&A...387..955G}, 
HL94: \citet{1994ApJ...437..281H}, 
Ho99: \citet{1999ApJS..124..195H}, 
Is84: \citet{Isaacman1984}, 
Li06: \citet{likkel_etal_2006}, 
LR96: \citet{1996ApJ...461..298L}, 
Lu01: \citet{Lumsden_etal_2001}, 
Ma16: \citet{2016ApJ...831L...3M}, 
ML15: \citet{MarquezLugo_etal_2015}, 
Ot13: \citet{2013ApJ...764...77O}, 
Ph83: \citet{1983MNRAS.203..977P}, 
Ph85: \citet{1985A&A...145..118P}, 
Ra93: \citet{1993MNRAS.263..695R}, 
RL08: \citet{2008AJ....135.1441R}, 
Ru01: \citet{2001AJ....121..362R}, 
Sm81: \citet{1981ApJ...244..835S}, 
St84: \citet{1984MNRAS.206..521S}, 
Vi99: \citet{1999A&A...342..823V}, 
We88: \citet{1988MNRAS.235..533W}.

\vspace{0.2cm}

\noindent \textbf{References for $T_\star$:}
Fr08: \citet{2008PhDT.......109F}, 
HH90: \citet{1990ApJ...353..200H}, 
Hu88: \citet{1988A&A...193..273H}, 
KJ89: \citet{1989ApJ...345..871K}, 
La00: \citet{2000ApJ...539..783L}, 
Lu01: \citet{Lumsden_etal_2001}, 
Ma15: \citet{2015ApJ...808..115M}, 
Ma16: \citet{2016ApJ...831L...3M}, 
Ot13: \citet{2013ApJ...764...77O}, 
Ot17: \citet{2017ApJ...838...71O}, 
Ph03: \citet{2003MNRAS.344..501P}, 
PM89: \citet{1989A&AS...81..309P}, 
PM91: \citet{1991A&AS...88..121P}, 
St02: \citet{2002ApJ...576..285S}, 
Sz09: \citet{2009ApJ...707L..32S}.

\vspace{0.2cm}

\noindent \textbf{References for $D$:}
Ak15: \citet{2015MNRAS.452.2911A}, 
BA91: \citet{1991A&A...250..165B}, 
BK18: \citet{2018MNRAS.480.1626B}, 
Be17: \citet{2017ApJ...837L..10B}, 
Cl14: \citet{2014A&A...569A..50C}, 
CS98: \citet{1998AJ....115.2466C}, 
CP00: \citet{2000ApJ...543..754C}, 
Da03: \citet{davis_etal_2003}, 
DM01: \citet{2001MNRAS.328..527D}, 
Fa15: \citet{2015MNRAS.452.2445F}, 
Fa18: \citet{Fang_etal_2018}, 
Fe97: \citet{1997ApJS..109..481F}, 
GH02: \citet{2002A&A...387..955G}, 
GM12: \citet{2012A&A...539A..47G}, 
Gu00: \citet{Guerrero_etal_2000}, 
Ha97: \citet{1997AJ....113.2147H}, 
He99: \citet{1999ApJ...517..782H}, 
HB04: \citet{2004ApJ...611..294H}, 
HL94: \citet{1994ApJ...437..281H}, 
Hs14: \citet{2014ApJ...787...25H}, 
Hy01: \citet{2001AJ....122..954H}, 
Kw10: \citet{2010ApJ...708...93K}, 
La16: \citet{2016ApJ...833..115L}, 
LF11: \citet{2011MNRAS.411.1395L}, 
Li06: \citet{likkel_etal_2006}, 
Lo93: \citet{1993IAUS..155..208L}, 
Lo01: \citet{1991A&A...241..526L}, 
ML15: \citet{MarquezLugo_etal_2015}, 
Ma98: \citet{1998A&A...329..683M}, 
Mi18: \citet{2019MNRAS.482..278M}, 
Mi97: \citet{1997MNRAS.288..777M},
Mi99: \citet{1999ApJ...520..714M}
Mo00: \citet{2000ApJ...537..853M}, 
OD03: \citet{2013AJ....145...92O}, 
Ot13: \citet{2013ApJ...764...77O}, 
Po09: \citet{2009A&A...502..189P}, 
Po09b: \citet{2009A&A...499..249P}, 
RL08: \citet{2008AJ....135.1441R}, 
RL12: \citet{2012MNRAS.423.3753R}, 
Sa11: \citet{2011AJ....141..134S}, 
Sh06: \citet{2006ApJS..167..201S}, 
Sh95: \citet{1995AJ....109.1173S}, 
St07: \citet{2007ApJ...671.1669S}, 
St08: \citet{2008ApJ...689..194S}, 
St16: \citet{2016ApJ...830...33S}, 
SS99: \citet{1999A&A...352..297S}, 
St84: \citet{1984MNRAS.206..521S}, 
Su04: \citet{2004A&A...421.1051S}, 
Sz11: \citet{2011MNRAS.416..715S}, 
Ty03: \citet{2003A&A...405..627T}, 
Va12: \citet{2012ApJ...751..116V}, 
Vi99: \citet{1999A&A...342..823V}, 
Wa18: \citet{2018A&A...620A.169W}, 
We88: \citet{1988MNRAS.235..533W}, 
Yu11: \citet{2011MNRAS.411.1035Y}.

\bsp 
\label{lastpage}
\end{document}